\documentclass[aps,pra,reprint,amsmath,amssymb,graphicx,longbibliography,superscriptaddress,nofootinbib]{revtex4-1}

\draft 

\usepackage{graphicx}
\usepackage{dcolumn}
\usepackage{bm}

\usepackage[utf8]{inputenc}
\usepackage[T1]{fontenc}
\usepackage{mathptmx}
\usepackage{dsfont}
\usepackage{xcolor}
\usepackage{todonotes}
\usepackage[unicode]{hyperref}
\hypersetup{
   unicode=true,          
   plainpages=false,
   colorlinks=true,       
   citecolor=blue,        
}

\usepackage{url}

\DeclareRobustCommand{\rev}[1]{#1}

\begin{document}


\title{Stochastic differential equation approach to understanding the population control bias in full configuration interaction quantum Monte Carlo}




\author{Joachim Brand}
\email{J.Brand@massey.ac.nz}
\affiliation{%
Dodd-Walls Centre for Photonic and Quantum Technologies, New Zealand Institute for Advanced Study and Centre for Theoretical Chemistry and Physics, Massey University, Auckland 0632, New Zealand%
}%

\author{Mingrui Yang}
\email{M.Yang4@massey.ac.nz}
\affiliation{%
Dodd-Walls Centre for Photonic and Quantum Technologies, New Zealand Institute for Advanced Study and Centre for Theoretical Chemistry and Physics, Massey University, Auckland 0632, New Zealand%
}%
\affiliation{%
MacDiarmid Institute for Advanced Materials and Nanotechnology, New Zealand Institute for Advanced Study and Centre for Theoretical Chemistry and Physics, Massey University, Auckland 0632, New Zealand%
}%

\author{Elke Pahl}%
\email{Elke.Pahl@auckland.ac.nz}
\affiliation{%
MacDiarmid Institute for Advanced Materials and Nanotechnology, Department of Physics, University of Auckland, Auckland 1010, New Zealand%
}%


\date{\today}

\begin{abstract}
We investigate a systematic statistical bias found in full configuration quantum Monte Carlo (FCIQMC) that originates from controlling a walker population with a fluctuating shift parameter. This bias can become the dominant error when the sign problem is absent, e.g.\ in bosonic systems. FCIQMC is a powerful statistical method for obtaining information about the ground state of a sparse and abstract matrix. We show that, when the sign problem is absent, the shift estimator has the nice property of providing an upper bound for the exact ground state energy and all projected energy estimators, while a variational estimator is still an upper bound to the exact energy with  substantially reduced bias.
A scalar model of the general FCIQMC population dynamics leads to an exactly solvable It\^o stochastic differential equation. It provides further insights into the nature of the bias and gives accurate analytical predictions for delayed cross-covariance and auto-covariance functions of the shift energy estimator and the walker number. The model provides a toe-hold on finding a cure for the population control bias. We provide evidence for non-universal power-law scaling  of the population control bias with walker number in the Bose-Hubbard model for various estimators of the ground state energy based on the shift or on projected energies. For the specific case of the non-interacting Bose-Hubbard Hamiltonian we obtain a full analytical prediction for the bias of the shift energy estimator.
\end{abstract}

\pacs{}

\maketitle 



%
%

%


\section{\label{sec:intro}Introduction}

Monte Carlo methods have the power to solve otherwise intractable computational problems by random sampling. They provide estimators for quantities of interest that will give the correct answer on average, but come with a statistical uncertainty, or Monte Carlo error. In addition to the statistical uncertainty, the estimators can also have a bias when the ensemble average of the estimator deviates from the exact value of the estimated quantity. Understanding and, if possible, removing such biases is an important challenge for Monte Carlo methods. In this paper we discuss the origin and nature of a systematic bias in the full configuration interaction quantum Monte Carlo (FCIQMC) method \cite{Booth2009}.

FCIQMC samples the ground state eigenvector of a quantum many-body Hamiltonian as is common for so-called projection Monte Carlo methods, which include diffusion Monte Carlo \cite{Anderson1975} and Greens function Monte Carlo \cite{Kalos1962}. FCIQMC provides access to statistical estimators for physical observables like the ground state energy. While the method generally applies to the computation of the dominant eigenvalue and eigenvector of an abstract square matrix as a stochastic variant of the power method \cite{enwiki:1002300566}, we will continue to use the language of quantum many-body physics where important applications lie. FCIQMC and its variations have been used with great success in quantum chemistry \cite{Cleland2012,Deustua2018}, the electronic structure of solid state systems \cite{Booth2013,Malone2016}, and ultra-cold atom physics \cite{Jeszenszki2020,Ebling2021,Yang2021a}

A particular feature of FCIQMC is that the detailed sign structure of the sampled coefficient vector is established spontaneously by the annihilation of walkers of opposite sign for large enough walker numbers \cite{Booth2009,Spencer2012}. Typically, the required walker number, known as the annihilation plateau, scales proportionally to the linear dimension of Hilbert space  \cite{Shepherd2014} and thus exponentially with the size of a physical system. This is a manifestation of the so-called sign problem, which is also present in other flavors of quantum Monte Carlo  \cite{Umrigar2007,Troyer2005,Iazzi2016}.
Much effort has gone into analyzing the sign problem for FCIQMC \cite{Kolodrubetz2013,Shepherd2014,Umrigar2015,Petras2021,Spencer2012} and into developing strategies and approximations for mitigating it \cite{Cleland2010,Gruneis2011,Tubman2016,Blunt2018,Ghanem2020,Ghanem2019,Blunt2019a}. Note that the sign problem is absent for real Hamiltonians with only non-positive off-diagonal matrix elements because no competition arises in assigning  signs of the coefficient vector elements. Such matrices are known as stoquastic matrices \cite{Bravyi2008}. The sign problem is equally absent from matrices obtained from a stoquastic matrix by flipping the signs of basis states.

Surprisingly little attention has been paid to a systematic statistical bias in the FCIQMC estimators that persists even for walker numbers above the annihilation plateau (or for FCIQMC with stoquastic matrices) \cite{Vigor2015}. This bias is known as the population control bias and is common to all known projection Monte Carlo methods that use population control \cite{Cerf1995}.
In FCIQMC the population is controlled by adjusting a scalar quantity known as the shift periodically during the simulation. After an initial equilibration period, the mean of the shift becomes an estimator for the exact ground state energy.
During a long history of study  \cite{Kalos1969,Hetherington1984,Umrigar1993,Cerf1995,Boninsegni2012}, it was concluded by several authors \cite{Hetherington1984,Umrigar1993,Cerf1995} that the population control bias in the eigenvector and in projected and growth estimators scales with $N_\mathrm{w}^{-1}$, where $N_\mathrm{w}$ is the number of walkers used in the calculation. A related result bounds the population control bias  proportional to $1/m$,
where $m$ is the number of sampled configurations (or non-zero coefficients of the stochastic representation of the ground state vector) at any one time \cite{Lim2015}.

Remarkably, there is numerical evidence contrary to the seeming consensus in the literature regarding the $N_\mathrm{w}^{-1}$ scaling. For diffusion Monte Carlo, power law decay of the bias with significantly slower decay exponents
was reported in Refs.~\cite{Boninsegni2012,Inack2018}.
In this paper
we report evidence for non-universal scaling of the population control bias with power law exponents as weak
as $\approx-0.4$ in FCIQMC data for Bose Hubbard chains with repulsive interactions. We further find a quadratic scaling of the population control bias \rev{in the total energy} with the size of the physical system \rev{when two-particle interactions are strong}. This is bad news for FCIQMC calculations on larger stoquastic Hamiltonians, where the population control bias becomes increasingly difficult to mitigate and poses a major challenge. Along similar lines, Ref.\ \cite{Inack2018} concluded that the numerical resources needed to retain a constant population control bias in diffusion Monte Carlo simulations of spin chains scale exponentially with system size.

Due to its growth with system size, the population control bias is particularly relevant for stoquastic Hamiltonians (typically found for bosonic problems) where it is the dominant systematic bias preventing accurate calculations of large scale quantum systems with limited memory resources (i.e.~limited number of walkers). For non-stoquastic Hamiltonians, on the other hand, the requirement to overcome the sign problem limits the system size and at the same time demands a minimum walker number such that typically regimes are accessed where the population control bias is so small that it is hard to detect in the presence of statistical errors, or of a larger systematic bias originating from the initiator approximation \cite{Cleland2010} when the latter is used. 

To mitigate the population control bias in projector Monte Carlo, it is possible to define formally unbiased estimators \cite{Hetherington1984,Nightingale1986,Umrigar1993}. The unbiased estimators can be obtained by reweighting the Monte Carlo time series data in post processing at the expense of additional stochastic errors. 
While this leads to an uncontrolled approximation, it has been shown to work well in practice in many cases  \cite{Nightingale1986,Nightingale1988,Umrigar1993,Vigor2015}.
An alternative strategy for suppressing the population control bias is to minimize the sampling noise with importance sampling.
Reference \cite{Inack2018a}  achieved this for Greens function Monte Carlo using a highly accurate neural network guiding function. Very recently, both reweighting and a simpler importance sampling scheme were combined to suppress the population control bias in FCIQMC \cite{Ghanem2021}.

In this work we derive exact relations for the population control bias in the shift and projected energy estimators. Projected energy estimators for the ground state energy are commonly used in projection Monte Carlo. We further
analyze the effect of noise in the FCIQMC algorithm in the framework of It\^o stochastic calculus \cite{Gardiner2009}. We assume that either the Hamiltonian is stoquastic, or the walker number is sufficiently large that the sign structure of the sampled coefficient vector is consistent with the exact eigenvector (i.e.~the walker number is above the annihilation plateau).  The main results are as follows:
\begin{itemize}
\item The shift estimator is an upper bound for the exact ground state energy and for all projected energy estimators, including the variational energy estimator, which is defined by a Rayleigh quotient.
\item We define a norm projected energy estimator. Excellent approximations to it are easy to compute from readily available walker number and shift data, and contain less bias than the shift estimator. While the norm projected energy is less biased than the shift estimator we find that the difference scales with $N_\mathrm{w}^{-1}$ in numerical data. The overall bias of the norm projected energy exhibits the same non-universal scaling as the shift estimator.
\item The variational energy estimator, which also provides an upper bound to the exact energy, is found to have a much reduced population control bias compared to the shift or norm projected energy estimators. We discuss an efficient way to calculate it numerically.
\item We derive an It\^o stochastic differential equation for the coupled dynamics of the walker number and the shift. A simplified scalar model can be solved exactly and  provides valuable insights into the role of the time step size and walker number control parameters. A particular prediction is that the population control bias in the energy estimator is independent of these parameters, which is confirmed by full numerical FCIQMC simulation results.
The analytic model also provides explicit formulas for the delayed auto- and cross-covariance functions of the walker number and the shift.
\item We analyze the reweighted estimators of Refs.~\cite{Hetherington1984,Nightingale1986,Umrigar1993} and find that they successfully remove most of the bias in our numerical examples. Unfortunately, the reweighting adds stochastic noise that grows in the limit of large reweighting depth where the population control bias is formally removed. Finding the optimal reweighting depth may require further research.
\item We derive explicit analytical expression for the population control bias in the non-interacting Bose-Hubbard chain.
We find that the population control bias is approximately extensive \rev{in the non-interacting case}, i.e. scales linearly with particle number, in contrast to the interacting system where quadratic scaling with system size was observed. Specifically for a single particle in the Hubbard chain the population control bias is asymptotically given by $2J/N_\mathrm{w}$ for large walker numbers $N_\mathrm{w}$, while it decays faster for small walker numbers. $J$ is the hopping parameter in the Hubbard chain.
\end{itemize}



This paper is organized as follows:
After introducing FCIQMC as a random process with its main equations in Sec.~\ref{sec:fciqmc}, and stating computational details in Sec.~\ref{sec:sim} we provide evidence for non-universal scaling laws of the population control bias in Sec.~\ref{sec:evidence}. Various energy estimators are defined and exact relations for the population control bias are derived  in Sec.~\ref{sec:exactpcb} before developing a scalar model that leads to a solvable stochastic differential equation in Sec.~\ref{sec:scalar}.
The reweighting procedure for unbiased estimators is derived and analyzed in the context of the present work in Sec.~\ref{sec:Reweighting}.
Section~\ref{sec:noiseinfciqmc} discusses explicit expressions linking the population control bias to the matrix structure of the Hamiltonian within the integer walker number FCIQMC algorithm. These are further applied to the non-interacting Bose-Hubbard chain,
before concluding in Sec.~\ref{sec:conclusion}.
Appendix \ref{sec:params} reports data on the influence of simulation parameters on the outcome and finds no significant dependence of the population control bias in the energy estimators on the forcing parameter of population control and the time step parameter.
A proof that the shift estimator is an upper bound for the projected energies is provided in
App.~\ref{sec:upperbound} and a detailed derivation of the exact solution of the stochastic differential equation with Greens functions in App.~\ref{sec:GreensFunctions}. Appendix \ref{sec:sampling} analyzes the detailed noise properties of the integer walker sampling algorithm and App.~\ref{sec:normSDE} derives a stochastic differential equation for the walker number in the sparse walker regime.



\section{Full FCIQMC equations} \label{sec:fciqmc}

The FCIQMC equations aim at sampling the ground state eigenvector of a matrix representation $\mathbf{H}$ of the quantum Hamiltonian. The algorithm is based on the iterative equations updating a coefficient vector $\mathbf{c}^{(n)}$
 and scalar shift $S^{(n)}$: 
\begin{align}
\label{eq:Cupdate}
\mathbf{c}^{(n+1)} & = [\mathds{1} + \delta\tau(S^{(n)}\mathds{1} - \check{\mathbf{H}})]\mathbf{c}^{(n)} ,\\
\label{eq:Supdate}
S^{(n+1)} & = S^{(n)}- \frac{\zeta}{\delta\tau}\ln\frac{N_\mathrm{w}^{(n+1)}}{N_\mathrm{w}^{(n)}}
- \frac{\xi}{\delta\tau}\ln\frac{N_\mathrm{w}^{(n+1)}}{N_\mathrm{t}} ,
\end{align}
where $\delta\tau$ is a time-step parameter. The walker number $N_\mathrm{w}^{(n)}$ will be discussed in more detail below together with its control parameters $N_\mathrm{t}$, $\zeta$, and $\xi$.

Equation \eqref{eq:Cupdate} performs, in an average sense, the projection by repeatedly multiplying the matrix $\mathbf{H}$ with the coefficient vector $\mathbf{c}^{(n)}$. If done exactly, it will suppress the norm of excited states exponentially in $n$ compared to the ground state. With the symbol $\check{\mathbf{H}}$ we indicate that deterministic matrix vector multiplication in Eq.~\eqref{eq:Cupdate} is replaced by a random process in FCIQMC.
In the original formulation with integer walker numbers \cite{Booth2009} this was achieved by a sequence of spawning, death and/or cloning steps for individual walkers. Modern variations of FCIQMC like the semistochastic version \cite{Petruzielo2012,Blunt2015} and fast randomized iteration algorithms \cite{Lim2015,Greene2019,Greene2020} modify the sampling procedures in order to reduce stochastic noise.

While the details of the sampling procedure do not matter for most parts of this work (they will be considered in App.~\ref{sec:sampling}), it is important that the sampling procedure is designed to achieve the correct vector-matrix multiplication on average, in the sense of an ensemble average for every single iteration step:
\begin{align} \label{eq:ensembleEV}
\mathrm{E}\left([\mathds{1} + \delta\tau(S^{(n)}\mathds{1} - \check{\mathbf{H}})]\mathbf{c}^{(n)} \right) &=[\mathds{1} + \delta\tau(S^{(n)}\mathds{1} - {\mathbf{H}})]\mathbf{c}^{(n)} ,
\end{align}
where $\mathrm{E}(\cdot)$ denotes the expected value of the sampling procedure for a given coefficient vector $\mathbf{c}^{(n)}$ and given $S^{(n)}$.

It is essential for the analysis in the rest of this work that we can
think of the sampling process as a multiplication with a noisy matrix. This justifies our notation, where $\check{\mathbf{H}}$ represents a random
matrix, which (ensemble) averages to the full matrix ${\mathbf{H}}$.
Note that this picture  may fail for non-stoquastic matrices when the walker number is too low to support sufficient walker annihilation. As a manifestation of the sign problem, effectively a different matrix is sampled on average in this case \cite{Spencer2012}. Thus we will assume in the following that  ${\mathbf{H}}$ is a stoquastic matrix (as will be true in all examples presented), or that the walker number is above the annihilation plateau.

Because Eq.~\eqref{eq:Cupdate} does not generally conserve the norm of the updated coefficient vector, it needs to be supplemented by a population control procedure, which is provided by Eq.~\eqref{eq:Supdate}.
The number of walkers $N_\mathrm{w}^{(n)}$ is computed from the coefficient vector by the 1-norm at each time step $n$
\begin{align}
N_\mathrm{w}^{(n)} =  \lVert\mathbf{c}^{(n)}\rVert_1\equiv \sum_i \left| c_i^{(n)} \right|  ,
\end{align}
where for now we assume that the elements of  the coefficient vectors $\mathbf{c}^{(n)}$ and the matrix $\mathbf{H}$ are real numbers.
The parameter $\zeta$ controls a damping term resisting the change in walker number whereas $\xi$ controls a restoring force that causes the walker number to eventually fluctuate around the pre-set target walker number $N_\mathrm{t}$.
The last term in Eq.~\eqref{eq:Supdate} was introduced in Ref.~\cite{Yang2020}, and the (more common) original walker control procedure  of Ref.~\cite{Booth2009} is recovered as the special case where $\xi =0$.

The dependence of the simulation results on the parameters $\zeta$ and $\xi$ was discussed in detail in Ref.~\cite{Yang2020}. In particular, the population control bias in the shift was found to be independent of the forcing parameter $\xi$ implying that the new walker control procedure of  Ref.~\cite{Yang2020} produces the same bias as the original one of Ref.~\cite{Booth2009}.  In App.~\ref{sec:params} we present further data showing no significant dependence of the population control bias of the shift and various projected energy estimators on either the forcing parameter $\xi$, or the time step size $\delta\tau$ even  in the presence of delayed update intervals. For this reason we set the forcing parameter to $\xi = \zeta^2/4$ in all numerical simulations in the main part of the paper, which corresponds to critical damping and produces optimal walker number control \cite{Yang2020}.
The full parameter dependence is however considered in the analytical derivations of Sec.~\ref{sec:scalar}, the results of which explain many findings of Ref.~\cite{Yang2020} including the insensitivity of the population control bias on the details of the population control procedure.

\section{Simulation details}\label{sec:sim}
Simulations were performed with the open source Julia package \texttt{Rimu.jl}  \cite{rimu2020}, written by the authors, and use the integer walker number FCIQMC algorithm of Ref.~\cite{Booth2009} supplemented with  the improved walker control protocol of Ref.~\cite{Yang2020}, as per Eqs.~\eqref{eq:Cupdate} and \eqref{eq:Supdate}. Energy estimators are computed as averages from a time series collected from the simulation discarding data from an initial equilibration phase.

\subsection{Estimating uncertainties}
Monte Carlo time series data is correlated over a finite time scale. In order to estimate the standard error, we remove these correlations by re-blocking \cite{Flyvbjerg1989} augmented by hypothesis testing to check that the correlations have been reduced to undetectable levels \cite{Jonsson2018}.

For energy estimators defined by a ratio of expected values, we separately calculate the sample means of the numerator and the denominator and treat them as correlated Gaussian variables, which should be true for a sufficiently long time series by virtue of the central limit theorem. The variances and the covariance of the sample means are estimated after re-blocking using the same number of blocking steps such that autocorrelations in both time series are below detection limit.
In a second step we determine the confidence interval of the corresponding ratio distribution with Monte Carlo error propagation using the package \texttt{MonteCarloMeasurements.jl}  \cite{bagge2020MCM.jl}. Throughout this paper (in plots) we report the median of the resulting distribution and error bars indicating the 68\% confidence interval (which is equivalent to a $1\sigma$ standard error for normally distributed random variables).

In general and unless explicitly noted we use long time series with $\Omega\sim 10^6$ Monte Carlo steps for the data analysis after allowing for an ample equilibration period of $\sim 10^5$ steps, independent of other parameters being varied in the same plot (e.g.~particle number $N$ or target walker number $N_\mathrm{t}$). This naturally leads to varying sizes of statistical error bars.

\subsection{Bose Hubbard Hamiltonian}
While most of the theoretical results presented in this work are independent of the specifics of the Hamiltonian,  all numerical FCIQMC simulations reported in this paper were done with the Bose Hubbard model \cite{Fisher1989} in  one spatial dimension with periodic boundary conditions (chain configuration) in real space. A total of $N$ bosonic particles can access $M$ lattice sites, which brings the dimension of Hilbert space to  $\binom{M+N-1}{N}$.
The model comprises on-site interaction between particles  characterized by a strength parameter $U$ and hopping to nearest neighbor sites described by the hopping strength $J>0$:
\begin{align}\label{eq:bhm}
H =  -J  \sum_{\langle i,j \rangle}\hat{a}_i^\dag \hat{a}_{j} + U \sum_i \hat{n}_i (\hat{n}_i -1).
\end{align}
Here ${\langle i,j \rangle}$ denotes that the summation is performed over all adjacent lattice sites. The operators $\hat{a}_{i}^\dag $ and $\hat{a}_{i}$ create and annihilate particles at sites $i$, respectively, and follow canonical bosonic commutation relations. The number operator $\hat{n}_i =\hat{a}_i^\dag \hat{a}_{i} $ counts the particles on lattice site $i$. For a one-dimensional chain of $M$ sites with periodic boundaries, the first summation consists of $2M$ terms, while the second one has $M$ terms.

The Bose Hubbard model is relevant to ultra-cold atom experiments  \cite{Greiner2002}, where readout at single atom level can be achieved with quantum gas microscopes \cite{Bakr2010}.

\section{Non-universal scaling of the population control bias}\label{sec:evidence}

\begin{figure}
\includegraphics[width=\columnwidth]{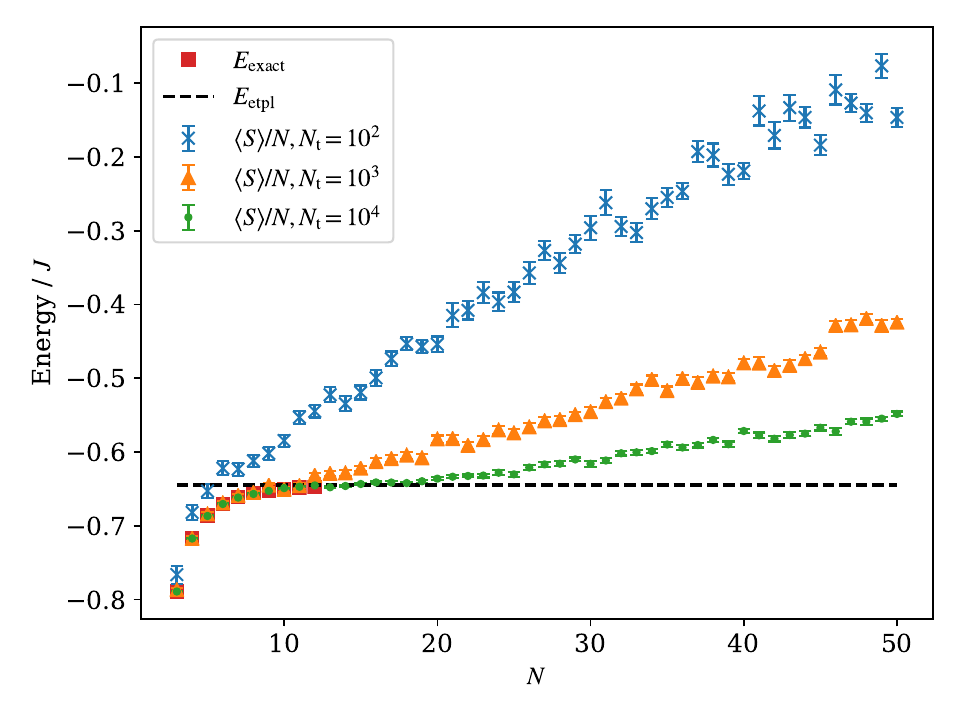}
 \caption{Population control bias vs.~system size.
Shown is the shift energy estimator for the energy per particle in a Hubbard chain with fixed filling factor of $M/N=1$ for FCIQMC calculations with different (fixed) values of the target walker number $N_\mathrm{t} \approx \langle N_\mathrm{w} \rangle$ as indicated. Also shown are exact diagonalization results for up to $N=12$ particles (red squares) and the extrapolated energy per particle for an infinite system (dashed black line). The deviation of the FCIQMC results from the exact (or extrapolated) values represents the population control bias per particle at the given walker number, which is seen to grow linearly with system size. FCIQMC calculations show data with up to $N=50$ particle with a Hilbert space dimension of  $\binom{M+N-1}{N} \approx 5 \times 10^{28}$.
We used $U/J = 6$, which lies in the Mott insulating regime and parameters $\zeta = 0.08$  at critical damping ($\xi = \zeta^2/4$) and $\delta\tau = 0.001 J^{-1}$.
 }
 \label{fig:pcb-2Nt}
\end{figure}

In this section we explore the scaling behavior of the population control bias with numerical results for the Bose Hubbard chain.
Figure \ref{fig:pcb-2Nt} serves to demonstrate how the population control bias becomes a real problem when scaling up the size of the physical system while being constrained with computer resources to work at fixed walker number. The walker number is an upper bound on the number of non-zero elements of the coefficient vector that have to be stored and thus provides an excellent proxy for the memory requirement. Figure \ref{fig:pcb-2Nt} shows estimators for the energy per particle for a Bose-Hubbard chain with one particle per lattice site as a function of the system size. As the energy per particle is an intensive quantity it is expected to become independent of particle number with deviations for small particle numbers due to finite-size effects. This is seen in the data from exact diagonalization (red squares) for up to 12 particles. The dashed line is the extrapolated energy per particle for the infinite chain.

The Monte Carlo data for the shift energy estimator is seen to lie above the exact values, which is a manifestation of the population control bias. The data also clearly suggests that the bias in the energy per particle at fixed walker number grows linearly with system size. It follows that the bias of the total energy grows quadratically with system size, and thus is not an extensive variable. This quadratic scaling with the number of particles suggests that the origin of the bias might be linked to the two-particle interactions present in the Hamiltonian.

The quantum state sampled in Fig.~\ref{fig:pcb-2Nt} corresponds to a Mott insulator state, which is characterized by small fluctuations of the number of particles per lattice site and a gap in the excitation spectrum (which opens for $U/J \gtrapprox 3.4$ \cite{Rossini2012}) as a consequence of the relatively high energy cost of having more than one boson on a given lattice site.

\begin{figure}
\includegraphics[width=\columnwidth]{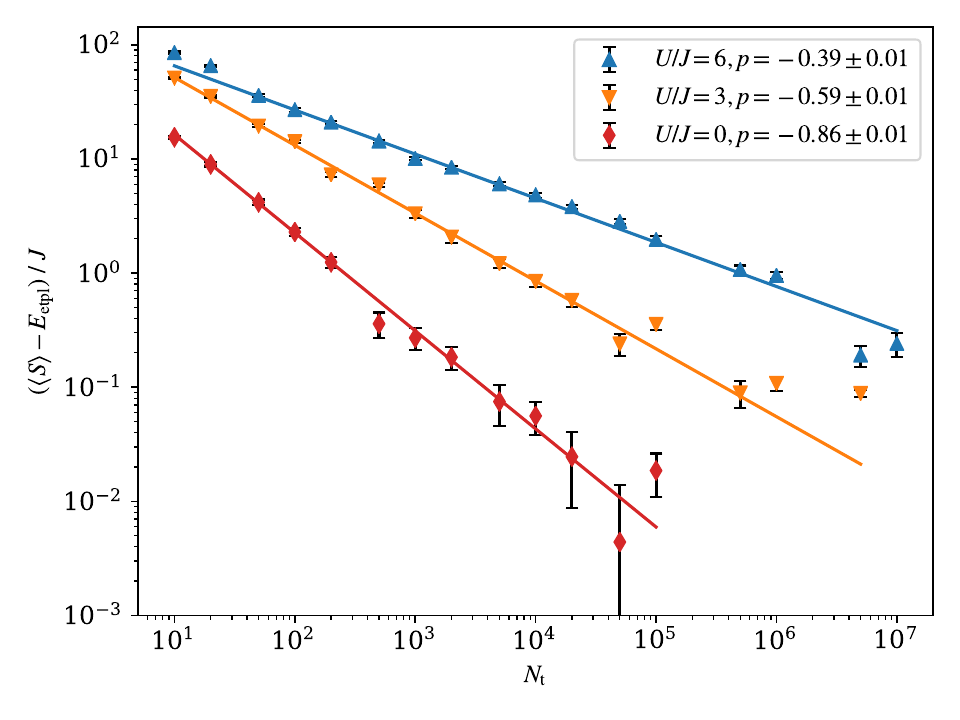}
 \caption{Non-universal power-law scaling of the population control bias with walker number. Shown is the shift estimator vs.~the  walker number for different interaction parameters in the real-space Bose-Hubbard chain with $N=50$ particles in $M=50$ lattice sites.  Lines are fits to the power law $\langle S\rangle = a+b \langle N_\mathrm{w}\rangle^p$, where the power $p$ varies significantly with the model parameters as indicated in the legend. Other parameters as in Fig.~\ref{fig:pcb-2Nt}.
 }
 \label{fig:pcb-scaling}
\end{figure}

Figure \ref{fig:pcb-scaling}  shows how the population control bias scales with the walker number. In the plots we report the target walker number $N_\mathrm{t}  \approx \langle N_\mathrm{w} \rangle$, since the fluctuations in the walker number are small when using the walker control procedure of Eq.~\ref{eq:Supdate} introduced in Ref.~\cite{Yang2020}. The data provides evidence that the bias scales as a simple power law $\sim N_\mathrm{t}^p$
over up to six decades for the strongly-interacting data at $U/J=6$. The power is also seen to depend strongly on the interaction parameter  $U/J$ defying the predictions of universal $N_\mathrm{t}^{-1}$ scaling in  Refs.~\cite{Hetherington1984,Umrigar1993,Cerf1995,Lim2015,Vigor2015}. In particular the strongly interacting Mott-insulating state presents a stubbornly slowly decaying population control bias.

\begin{figure}
\includegraphics[width=\columnwidth]{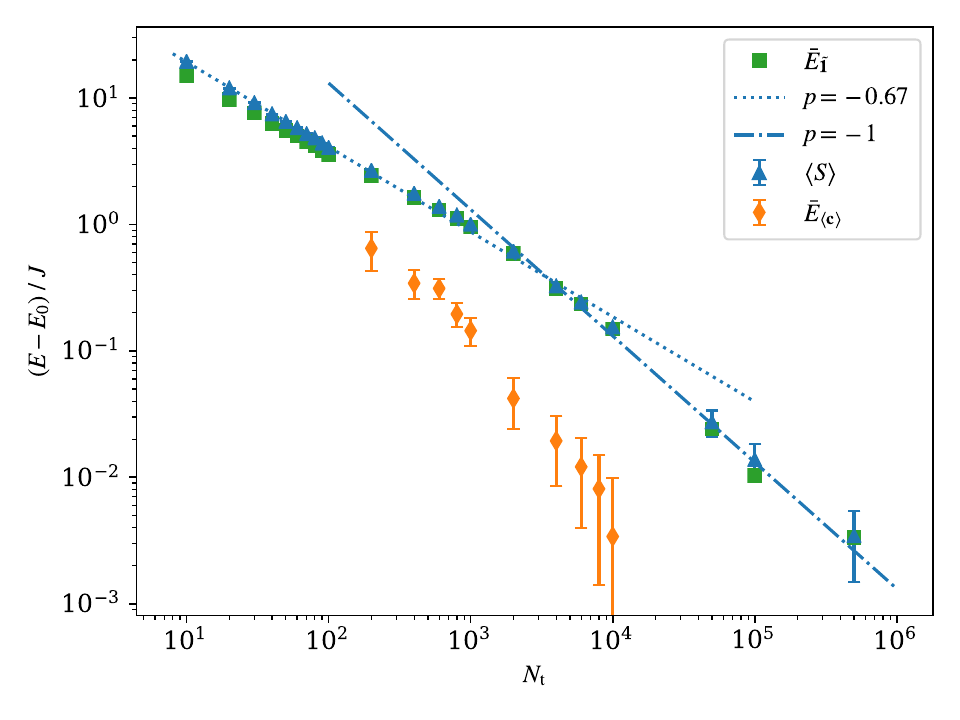}
 \caption{Crossover power laws for different energy estimators. Shown are the shift (blue triangles), norm projected energy (green squares), and variational energy estimators (orange diamonds) vs.~walker number for the real-space Bose-Hubbard chain with $N=20$ particles in $M=20$ lattice sites and $U/J =6$.
For the shift and projected energy estimators two different regimes can be distinguished that follow an approximate power-law behavior. Lines are power-law fits to the corresponding subsets of data with exponents as indicated.
Parameters of the calculation are $\zeta = 0.08$  at critical damping ($\xi = \zeta^2/4$) and $\delta\tau = 0.001 J^{-1}$ with $\Omega=4\times10^6$ time steps. The dimension of Hilbert space is $\binom{M+N-1}{N} \approx 7 \times 10^{10}$. \rev{$E_0/J = -12.894$ was estimated from a calculation with $N_\mathrm{t}=10^7$ walkers.} }
 \label{fig:variationalenergy}
\end{figure}

We note that we do find $N_\mathrm{t}^{-1}$ scaling consistently in smaller systems, e.g. for the $N=M=10$ Hubbard chain even in the Mott-insulating regime. The dimension of Hilbert space in this case is $\approx 10^5$, which is much smaller than the system of Fig.~\ref{fig:pcb-scaling}. 

Figure \ref{fig:variationalenergy} shows an interesting intermediate case with  $N=M=20$ where the dimension of Hilbert space is $\approx 10^{11}$. Here we observe a crossover between two regimes with slow power-law scaling for small walker number, and $N_\mathrm{t}^{-1}$ scaling for $N_\mathrm{t} \gtrsim 10^4$.
In addition to the shift energy estimator, Fig.~\ref{fig:variationalenergy} also shows the norm projected energy $\bar{E}_{\tilde{\mathbf{1}}}$, which follows the shift very closely for this example, and the variational energy estimator $\bar{E}_{\langle\mathbf{c}\rangle}$, which has a smaller bias. Both estimators will be defined in Sec.~\ref{sec:exactpcb}, where theoretical arguments regarding their scaling with walker number will be presented.

\section{Exact relations for the population control bias}\label{sec:exactpcb}

We  consider the steady-state limit of the FCIQMC equations \eqref{eq:Cupdate} and \eqref{eq:Supdate}, where the coefficient vector $\mathbf{c}^{(n)}$ and the shift $S^{(n)}$ will be fluctuating around some expected value (obtained as an ensemble average over noise/random number realizations), which is identical to the long-time average.
Let $\langle \cdot \rangle$ denote the long time average over a stationary time series.
We consider the averages for the FCIQMC equations and start with the shift update equation \eqref{eq:Supdate}.
Noting that the time-series average is translationally invariant in the steady state and thus
\begin{align}
\left\langle S^{(n+1)} \right\rangle &= \left\langle S^{(n)} \right\rangle \equiv \left\langle S \right\rangle , \\
\left\langle  \ln N_\mathrm{w}^{(n+1)} \right\rangle &= \left\langle \ln N_\mathrm{w}^{(n)} \right\rangle
\equiv \left\langle \ln{N_\mathrm{w}} \right\rangle ,
\end{align}
most terms cancel. We obtain
\begin{align}
 0 = \left\langle \ln \frac{N_\mathrm{w}^{(n)}}{N_\mathrm{t}} \right\rangle =
\left\langle \ln{N_\mathrm{w}} \right\rangle - \ln {N_\mathrm{t}} ,
\end{align}
or that the logarithm of the walker number averages to the logarithm of the target walker number $\ln N_\mathrm{t}$.

Averaging the coefficient vector update equation \eqref{eq:Cupdate} yields
\begin{align} \label{eq:prodave}
0 =  \left\langle S^{(n)} \mathbf{c}^{(n)} \right\rangle -  \left\langle \check{\mathbf{H}}\mathbf{c}^{(n)} \right\rangle .
\end{align}
The second term on the right hand side can be simplified further.
Because the long-time limit is equivalent to (or implies) an ensemble average over sampling noise, we may take the latter before the former. Equation\ \eqref{eq:ensembleEV} thus implies
\begin{align} \label{eq:prodH}
 \left\langle \check{\mathbf{H}}\mathbf{c}^{(n)} \right\rangle =
 \left\langle \check{\mathbf{H}} \right\rangle \left\langle \mathbf{c^{(n)}}\right\rangle \equiv
 \mathbf{H} \left\langle \mathbf{c}\right\rangle .
\end{align}
This is the same result that we would have obtained by treating $\check{\mathbf{H}}$ as a random matrix with independent random numbers that are uncorrelated with the fluctuations in the time series $\mathbf{c^{(n)}}$. From now on we will thus assume that this is the case, as it simplifies the analysis.

The first term in Eq.~\eqref{eq:prodave}, however, is a product of fluctuating variables, which are not independent and therefore
\begin{align}\label{eq:Sc_cov}
 \left\langle S^{(n)} \mathbf{c}^{(n)} \right\rangle =  \left\langle S \right\rangle  \left\langle \mathbf{c} \right\rangle + \operatorname{cov}( S^{(n)}, \mathbf{c}^{(n)}) ,
\end{align}
where we define the covariance as
\begin{align}
\operatorname{cov}(a,b) =  \left\langle (a - \left\langle a \right\rangle)(b -  \left\langle b \right\rangle)  \right\rangle .
\end{align}
Note that the covariance between a scalar and a vector is to be taken elementwise on the vector. The final result for the averaged equation for the coefficient update is
\begin{align} \label{eq:covSC}
 \left\langle S \right\rangle  \left\langle \mathbf{c} \right\rangle -  \mathbf{H} \left\langle \mathbf{c}\right\rangle = -
 \operatorname{cov}( S^{(n)},  \mathbf{c}^{(n)}) .
\end{align}
For vanishing covariance we re-cover the time-independent Schr\"odinger equation (or eigenvalue equation).
The fact that fluctuations of the shift and the coefficient vector are coupled gives rise to the population control bias as we will see in the following sections.

\subsection{Projected energy estimator}\label{sec:pEst}

Projected energy estimators are commonly used in FCIQMC and other projector Monte Carlo methods. For an arbitrary vector $\mathbf{y}$ we define the projected energy by
\begin{align} \label{eq:p_energy}
\bar{E}_\mathbf{y} \equiv \frac{ \left\langle \mathbf{y}^\dag \mathbf{H} \mathbf{c} \right\rangle }{\left \langle \mathbf{y}^\dag\mathbf{c} \right\rangle} =
\frac{ \mathbf{y}^\dag \mathbf{H} \left\langle\mathbf{c} \right\rangle }{ \mathbf{y}^\dag \left\langle\mathbf{c} \right\rangle} ,
\end{align}
where $\mathbf{a}^\dagger \mathbf{b}$ is the scalar product of two (column) vectors.
When the coefficient vector samples the exact eigenstate it will yield the exact ground state energy if $\mathbf{y}$ has non-negligible overlap with the eigenvector. The quantity $\bar{E}_\mathbf{y}$ is easy to compute and can provide low fluctuations if a good choice of $\mathbf{y}$ can be found.
We can easily derive the following relation to the shift estimator from Eq.~\eqref{eq:covSC} by projection with $\mathbf{y}^\dag$ from the left
\begin{align} \label{eq:pEcov}
\left\langle S \right\rangle  - \bar{E}_\mathbf{y} =  -\frac{
 \operatorname{cov}( S^{(n)}, \mathbf{y}^\dag\mathbf{c}^{(n)})}{ \left\langle \mathbf{y}^\dag  \mathbf{c} \right\rangle} \ge 0.
\end{align}
Although this equation has no direct information about the  population control bias in either the shift or  the projected energy, it may still be useful by the fact that a difference between the average shift and the projected energy indicates the presence of a non-negligible population control bias. The quality of the projected energy estimator depends on both the quality of the sampled coefficient vector $\left\langle\mathbf{c} \right\rangle$ and the quality of the vector $\mathbf{y}$. Clearly, if the exact eigenvector is chosen for $\mathbf{y}$, or a good approximation of it, the projected energy $\bar{E}_\mathbf{y}$ can be made arbitrarily close to the exact energy, even when the quality of the sampled coefficient vector is poor.

The inequality in Eq.~\eqref{eq:pEcov} requires a separate proof, which is provided in App.~\ref{sec:upperbound} using methods of Sec.~\ref{sec:scalar}. It states that the shift energy estimator is an upper bound on the projected energy.
This is a powerful result, because it is true for arbitrary choices of the vector $\mathbf{y}$. We explore the consequences for specific choices of $\mathbf{y}$ in the following.

\subsection{Shift estimator}

We can obtain an  explicit expression for the population control bias of the shift energy estimator by substituting  the ground state eigenvector $\mathbf{c}_0$ for  the vector $\mathbf{y}$ in Eq.\ \eqref{eq:pEcov} to obtain
\begin{align} \label{eq:pcb}
 \left\langle S \right\rangle  -  E_0 = -\frac{
 \operatorname{cov}( S^{(n)},  \mathbf{c}_0^\dagger\mathbf{c}^{(n)})}{ \left\langle \mathbf{c}_0^\dagger \mathbf{c} \right\rangle}
 \ge 0,
\end{align}
where $E_0$ is the exact ground state energy. The right hand side of the equation provides an exact expression for the population control bias in the shift estimator. The inequality further assures that the shift estimator is an upper bound for the exact ground state energy.
The covariance expression is important conceptually, as it indicates how the coupled fluctuations in the shift and projected coefficient vector cause the population control bias.
From the properties of the covariance we can also obtain an upper bound
\begin{align}
 \left\langle S \right\rangle  -  E_0 \le \sqrt{\operatorname{var}(S) \frac{\operatorname{var}\left( \mathbf{c}_0^\dagger \mathbf{c} \right)}{\left\langle \mathbf{c}_0^\dagger \mathbf{c} \right\rangle^2}} ,
\end{align}
which indicates that reducing the fluctuations of both the shift and the coefficient vector is an effective strategy to suppress the population control bias.

\subsection{Norm projected energy estimator} \label{sec:normprojected}

As another special case let us consider the choice $\mathbf{y} = \tilde{\mathbf{1}}$, where we define the vector $\tilde{\mathbf{1}}$ to have entries of modulus 1 that carry the sign of the exact eigenvector  $ \mathbf{c}_0$. We have already committed ourselves to the case where the walker number is above the minimum required to mitigate the sign problem, and thus can
further assume that the sign structure of the fluctuating vector $\mathbf{c}^{(n)}$ is consistent with that of the exact eigenvector. 
The overlap with the coefficient vector thus produces the one-norm $\tilde{\mathbf{1}}^\dag \mathbf{c}^{(n)} = \lVert \mathbf{c}^{(n)}\rVert_1 = N_w^{(n)} $. Hence we obtain from Eq.~\eqref{eq:pEcov}
\begin{align} \label{eq:nEnergy}
\left\langle S \right\rangle  - \bar{E}_{\tilde{\mathbf{1}}} =  -\frac{
 \operatorname{cov}( S^{(n)},  N_\mathrm{w}^{(n)})}{ \left\langle  N_\mathrm{w} \right\rangle} \ge 0,
\end{align}
where the norm projected energy estimator is
\begin{align} \label{eq:E1tilde}
 \bar{E}_{\tilde{\mathbf{1}}} =   \frac{ \left\langle {\tilde{\mathbf{1}}}^\dag \mathbf{H} \mathbf{c} \right\rangle }{\left\langle  N_\mathrm{w} \right\rangle} .
\end{align}
Thus, the shift estimator is an upper bound for the norm projected energy.
The advantage of the norm projected energy estimator is that it can be easily calculated from Eq.\ \eqref{eq:nEnergy} using only shift and walker number data, which is collected anyway and thus does not require additional computational load at run time.

An approximation to the norm projected energy can also be obtained from averaging instantaneous time series data. This can be convenient for practical reasons.
We define
\begin{align}\label{eq:growth_witness_biased}
G^{(n)} &= S^{(n)} - \frac{N_\mathrm{w}^{(n+1)} - N_\mathrm{w}^{(n)}}{\delta \tau N_\mathrm{w}^{(n)}} ,
\end{align}
and call the average $\langle G \rangle$ the growth estimator.
It is easy to show that the average growth estimator evaluates to
\begin{align}\label{eq:growth_est_biased}
\langle G \rangle = \left\langle \frac{{\tilde{\mathbf{1}}}^\dag \mathbf{H} \mathbf{c}^{(n)}}{  N_\mathrm{w}^{(n)}} \right\rangle .
\end{align}
The growth estimator becomes equivalent to the norm projected energy estimator  $\langle G \rangle =  \bar{E}_{\tilde{\mathbf{1}}}$ for infinite time series averages. For finite averages it is still a good approximation
due to the fact that the walker number $N_\mathrm{w}^{(n)}$ in the denominator does not fluctuate strongly. In fact, the fluctuations in the walker number can be controlled by the parameters $\zeta$ and $\xi$, as discussed in Ref.~\cite{Yang2020}, and are typically sub-Poissonian, i.e. $\operatorname{var}(N_\mathrm{w}) \ll \langle N_\mathrm{w}\rangle$.

\rev{
Note that the quantity on the right hand side of Eq.~\eqref{eq:growth_est_biased} has another interpretation as the weighted average of the ``local energy''. Indeed, assuming a trivial sign structure (absence of the sign problem), the row vector $\tilde{\mathbf{1}}^\dag \mathbf{H}$ represents the column sum of the Hamiltonian also known as the local energy \cite{Umrigar1993,Kalos2007}.
}

\begin{figure}
\includegraphics[width=\columnwidth]{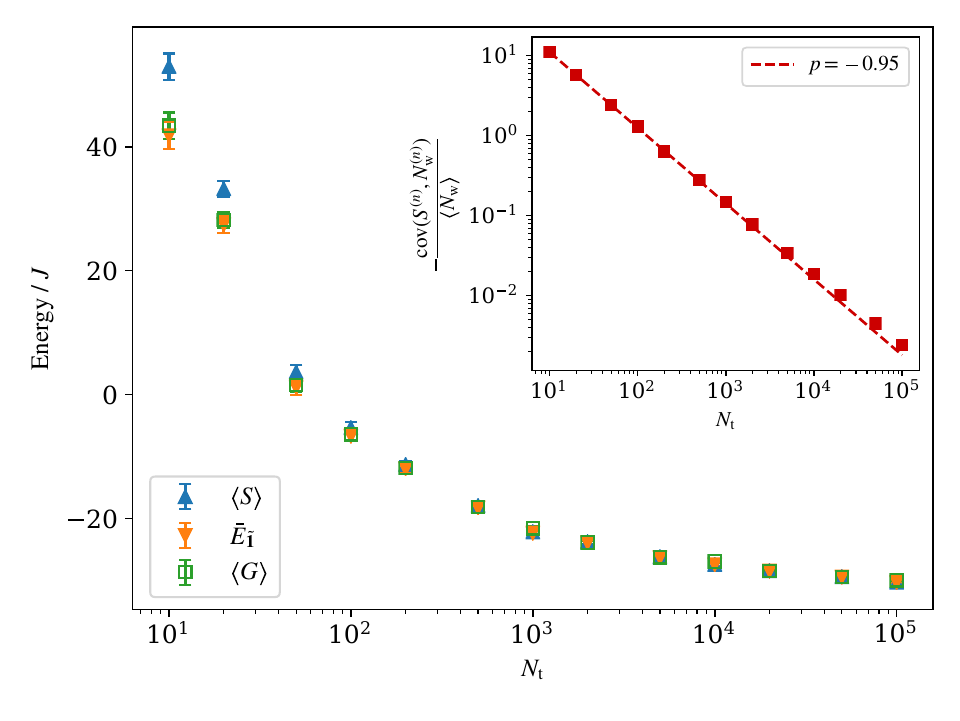}
 \caption{Energy estimators for the Bose Hubbard chain with $M=N=50$ and $U/J=6$.
 The inset shows the difference of the shift and the norm projected energy (by the right hand side of Eq.~\eqref{eq:nEnergy}) on a doubly logarithmic scale. The dashed line is a power-law fit with exponent close to $-1$. Parameters as in Fig.~\ref{fig:pcb-scaling}.
  }
 \label{fig:normprojectedenergy}
\end{figure}

Figure \ref{fig:normprojectedenergy} shows the norm projected energy and the growth estimator together with the shift  for the Bose Hubbard chain with $M=N=50$ and $U/J=6$. It is seen that the projected and the growth estimators essentially agree, and have less bias than the shift estimator for small walker numbers. Asymptotically, however they show the same scaling for large $N_\mathrm{w}$. The inset shows the difference of the shift and the norm projected energy (by the right hand side of Eq.~\eqref{eq:nEnergy}) on a doubly logarithmic scale. We find that this difference exhibits $1/N_\mathrm{w}$ scaling, which explains why asymptotically both the shift and the norm projected energy show a population control bias with the same slower-than-$1/N_\mathrm{w}$ scaling.

\subsection{Variational energy estimator}

Another interesting case is the projection onto the averaged vector $\mathbf{y} = \langle\mathbf{c}\rangle$. In this case the  energy estimator becomes the Rayleigh quotient
\begin{align} \label{eq:Rayleigh}
\bar{E}_{\langle\mathbf{c}\rangle} = \frac{\langle\mathbf{c}\rangle^\dag \mathbf{H} \langle\mathbf{c}\rangle }{\langle\mathbf{c}\rangle^\dag\langle\mathbf{c}\rangle} ,
\end{align}
which, by the variational theorem, provides an upper bound to the exact ground state energy $\bar{E}_{\langle\mathbf{c}\rangle} \ge E_0$.
For the difference from the average shift we obtain the relation
\begin{align}
\left\langle S \right\rangle  - \bar{E}_{\langle\mathbf{c}\rangle}  =  -\frac{
 \operatorname{cov}\left( S^{(n)}, {\langle\mathbf{c}\rangle}^\dag\mathbf{c}^{(n)}\right)}{{\langle\mathbf{c}\rangle} ^\dag  {\langle\mathbf{c}\rangle} } \ge 0.
\end{align}
Comparing this expression to the exact expression for the population control bias of Eq.~\eqref{eq:pcb}, we expect that the variational estimator $\bar{E}_{\langle\mathbf{c}\rangle}$ will be a much better estimator for the exact ground state energy than the averaged shift $\left\langle S \right\rangle$, because most of the population control bias has already been removed.
A strong reduction of the population control bias in the variational energy compared to the shift and projected energy estimators can be seen in Fig.~\ref{fig:variationalenergy} in calculations with the Bose Hubbard model with $N=20$ particles in $M=20$ sites.

\subsubsection*{Scaling with walker number}

In order to quantify how the population control bias in the variational and the norm projected energy estimators scale with walker number it is useful to introduce the
difference between the exact and the averaged eigenvector $\delta {\langle\mathbf{c}\rangle} = {\langle\mathbf{c}\rangle} -  \mathbf{c}_0$.
From the definition of the variational energy \eqref{eq:Rayleigh} we obtain
\begin{align}
\bar{E}_{\langle\mathbf{c}\rangle} - E_0
 & = \frac{\delta\langle\mathbf{c}\rangle^\dag ( \mathbf{H} - E_0 \mathds{1}) \delta \langle\mathbf{c}\rangle }{\langle\mathbf{c}\rangle^\dag\langle\mathbf{c}\rangle}  \ge 0 ,
\end{align}
where the inequality on the right holds because $\mathbf{H} - E_0 \mathds{1}$ is a positive semidefinite matrix.
The denominator can be bounded by the square of the walker number
\begin{align}
{\langle\mathbf{c}\rangle^\dag\langle\mathbf{c}\rangle} = \sum_i \langle c_i \rangle^2 \le \left\langle \sum_i |c_i| \right\rangle^2
= \langle N_\mathrm{w} \rangle^2 ,
\end{align}
where the equality will hold if the coefficient vector has a single nonzero element, i.e\ all walkers congregate on a single configuration.
Thus we obtain a lower bound for the bias in the variational energy estimator
\begin{align} \label{eq:boundVEloose}
\bar{E}_{\langle\mathbf{c}\rangle} - E_0 & \ge \frac{\delta\langle\mathbf{c}\rangle^\dag ( \mathbf{H} - E_0 \mathds{1}) \delta \langle\mathbf{c}\rangle }{\langle N_\mathrm{w} \rangle^2} \ge 0  .
\end{align}
A similar procedure can be performed for the norm-projected estimator to obtain
\begin{align} \label{eq:normPEdeltac}
\bar{E}_{\tilde{\mathbf{1}}} - E_0 &=
 \frac{ {\tilde{\mathbf{1}}}^\dag( \mathbf{H} - E_0 \mathds{1}) \delta\left\langle \mathbf{c} \right\rangle }{\langle N_\mathrm{w} \rangle} .
\end{align}
We can now reason about the scaling with walker number. When the norm projected energy estimator exhibits power law decay $\sim N_\mathrm{w}^{-\alpha}$ we may expect a faster decay for the variational energy estimator of Eq.~\eqref{eq:boundVEloose} with $\sim N_\mathrm{w}^{-2\alpha}$ because of the squared appearance of the walker-number dependent quantities ${\delta\left\langle \mathbf{c} \right\rangle }/ {\langle N_\mathrm{w} \rangle}$.

Note that Eq.~\eqref{eq:boundVEloose} provides a lower bound, which will be a good estimate only when the coefficient vector is highly concentrated on one or a few nonzero elements where stochastic noise will be small and the population control bias will not be a big issue anyway. A more interesting regime is the sparse walker limit where the number of non-zero coefficients in the exact eigenvector is much larger than the available number of walkers. In this regime we can derive a tighter bound by assuming that the
average number of walkers on each configuration is smaller than unity and thus
\begin{align}
{\langle\mathbf{c}\rangle^\dag\langle\mathbf{c}\rangle} = \sum_i \langle c_i \rangle^2 < \left\langle \sum_i |c_i| \right\rangle
= \langle N_\mathrm{w} \rangle .
\end{align}
This leads to a revised lower bound for the bias in the variational energy estimator
\begin{align} \label{eq:boundVE}
\bar{E}_{\langle\mathbf{c}\rangle} - E_0 & > \frac{\delta\langle\mathbf{c}\rangle^\dag ( \mathbf{H} - E_0 \mathds{1}) \delta \langle\mathbf{c}\rangle }{\langle N_\mathrm{w} \rangle} \quad\textrm{(sparse walkers)} .
\end{align}
Now, the bias in the variational energy estimator is bounded from below to a more slowly decaying power law due to the changed exponent of the walker number in the denominator.
In cases where the bias scales as $N_\mathrm{w} ^{-1}$ overall the numerator on the right hand side of Eq.~\eqref{eq:normPEdeltac} must be independent of walker number. We can thus expect the corresponding numerator  in Eq.~\eqref{eq:boundVE} to be constant as well.
For the variational energy bias this means that  it is bounded from below by $N_\mathrm{w} ^{-1}$ and thus it cannot decay any faster.
In the situation of Sec.~\ref{sec:evidence} where the norm projected energy has a bias that decays with a slower power law $\sim N_\mathrm{w}^{-\alpha}$ with $\alpha <1$, the situation is worse because the lower bound for the bias in the variational energy bias of Eq.~\eqref{eq:boundVE} decays even more slowly with $N_\mathrm{w}^{1 - 2 \alpha}$.

Note that the sparse walker regime is not the asymptotic regime for large walker number where, for any finite-sized matrix, the number of walkers on an individual configuration will eventually become larger than one.

%

\subsubsection*{Computation via replica trick}
In order to compute the variational energy estimator without keeping a long-time average of the whole state-vector around we use the replica trick to propagate two statistically independent fluctuating state vectors $\mathbf{c}_a^{(n)}$ and $\mathbf{c}_b^{(n)}$ with $\operatorname{cov}(\mathbf{c}_a^{(n)},\mathbf{c}_b^{(n)}) = 0$ and $\langle \mathbf{c}^{(n)}_a \rangle = \langle \mathbf{c}^{(n)}_b \rangle =\langle \mathbf{c} \rangle$. In Ref.~\cite{Overy2014} the replica trick was used to sample single and two-particle reduced density matrices from which the numerator and denominator of the Rayleigh quotient \eqref{eq:Rayleigh} can be obtained. Here we show how the variational energy estimator can be  calculated directly without the intermediate sampling of reduced density matrices.

We denote the finite sample mean of a fluctuating quantity $X^{(n)}$ by $\overline{X} = \Omega^{-1} \sum_n X^{(n)}$, where $\Omega$ is the sample size.
The numerator of the variational energy of Eq.~\eqref{eq:Rayleigh} can be obtained as the limit of samples means
\begin{align}
{\langle\mathbf{c}\rangle^\dag \mathbf{H} \langle\mathbf{c}\rangle } &= \lim_{\Omega\to\infty} {\overline{\mathbf{c}_a^\dag \mathbf{H} \mathbf{c}_b}} 
=\lim_{\Omega\to\infty}  {\overline{S_a \mathbf{c}_a^\dag \mathbf{c}_b}} =\lim_{\Omega\to\infty}  {\overline{S_b \mathbf{c}_a^\dag \mathbf{c}_b}} ,
\end{align}
where the last two equalities follow from Eqs.~\eqref{eq:prodave} and \eqref{eq:prodH}.  
Evaluating the sample mean with the full Hamiltonian matrix yields a smaller variance than using the expressions with the shift, but this  does not result in a significant difference in the standard error. Using the shift expressions instead avoids calculating overlaps with the Hamiltonian matrix and saves a significant amount of computer time.  
Averaging the last two expressions yields better statistics than taking each individually. We thus define the variational estimator of  replicas $a$ and $b$ by
\begin{align} \label{eq:Evab}
\bar{E}_{\mathrm{v}ab} & =  \frac{\overline{(S_a+S_b) \mathbf{c}_a^\dag \mathbf{c}_b}}{2\overline{\mathbf{c}_a^\dag \mathbf{c}_b}} ,
\end{align}
which only requires evaluating and storing  a time series of dot products of the instantaneous coefficient vector replicas $(\mathbf{c}_a^{(n)})^\dag \mathbf{c}_b^{(n)}$ along with the value of the shift for each replica. This is efficient in distributed calculations where the required parts of the coefficient vectors reside on the same computer node. It also requires much less storage and computer time than sampling reduced density matrices.

Obtaining this variational estimator becomes difficult in regimes of small walker number and large Hilbert space size because the dot products of the sparse and statistically independent coefficient vectors  will be zero for most time steps. This leads to a small and wildly fluctuating denominator in Eq.~\eqref{eq:Evab}, which makes the distribution of the ratio ill behaved. The statistics of the variational estimator can be vastly improved in this case by propagating more than two replicas at the same time, and obtaining the variational estimator after averaging the denominator and numerator separately over pairs of replicas. The finite sample variational estimator then becomes 
\begin{align} \label{eq:Ev}
\bar{E}_\mathrm{v} & =  \frac{\sum_{a<b}^R \overline{(S_a+S_b) \mathbf{c}_a^\dag \mathbf{c}_b}}{2\sum_{a<b}^R \overline{\mathbf{c}_a^\dag \mathbf{c}_b}} ,
\end{align}
where the sums run over distinct pairs out of the $R$ replicas.
Clearly, the variational energy of Eq.~\eqref{eq:Rayleigh} is obtained in the limit of large sample size:
\begin{align}
\bar{E}_{\langle\mathbf{c}\rangle} =  \lim_{\Omega\to\infty} \bar{E}_{v}.
\end{align}

Adding more replicas has been found more efficient in reducing the standard error of the denominator in Eq.~\eqref{eq:Ev} compared to increasing the number of time steps. Indeed, we have found that the dot products of the coefficient vectors are only weakly correlated between different combinations of replicas. Neglecting such correlations, the standard deviation of the denominator in Eq.~\eqref{eq:Ev} then scales $\propto 1/\sqrt{R(R-1) \Omega}$, where $R$ is the number of replicas and $\Omega$ is the number of time steps taken. As long as the computational cost of evaluating dot products is negligible, the overall computational cost for replica calculations will scale with $R \Omega$. Thus increasing $R$ leads to a better ratio of effect to cost than increasing $\Omega$. This will change for large $R$ when the quadratically growing computational cost of evaluating the dot products dominates and thus increasing $R$ further brings no relative advantage over increasing $\Omega$.

The variational energies shown in Fig.~\ref{fig:variationalenergy} were obtained from Eq.~\eqref{eq:Ev} using $R=3$ replicas, which dramatically reduced the fluctuations compared to a previous calculation with only two replicas. The increase in CPU time compared to a single replica calculation is still approximately given by the number of replicas, i.e.\ three in this case. On the other hand it can be seen in Fig.~\ref{fig:variationalenergy} that the reduction of the population control bias by using the variational estimator instead of the shift estimator is approximately equivalent to increasing the walker number by an order of magnitude.

\section{A scalar model} \label{sec:scalar}

More insight into the causes and manifestations of the population control bias can be gained by considering the effect of noise injected by the FCIQMC random sampling procedure on the dynamics of scalar quantities. Here we consider the walker number $N_w^{(n)} =  \tilde{\mathbf{1}}^\dag \mathbf{c}^{(n)}$ obtained by projection of the coefficient vector with the $\tilde{\mathbf{1}}$ vector defined in Sec.~\ref{sec:normprojected}. A generalized procedure is used in App.~\ref{sec:upperbound} to obtain the inequality of Eq.~\eqref{eq:pEcov}.

\subsection{A stochastic difference equation} \label{sec:scalardiffeq}

Projecting the FCIQMC equation \eqref{eq:Cupdate} for the coefficient vector from the left with $\tilde{\mathbf{1}}^\dagger$ yields a scalar equation for the particle number
\begin{align} \label{eq:scalarNorm}
N_\mathrm{w}^{(n+1)} = N_\mathrm{w}^{(n)} + \delta\tau\left(S^{(n)}N_\mathrm{w}^{(n)}
- \tilde{\mathbf{1}}^\dagger\check{\mathbf{H}} \mathbf{c}^{(n)}\right) .
\end{align}
To make further progress, we separate the right hand side into a deterministic part
and a noise part:
\begin{align}
\nonumber
N_\mathrm{w}^{(n+1)} &= N_\mathrm{w}^{(n)} + \left(S^{(n)}N_\mathrm{w}^{(n)}
- \tilde{\mathbf{1}}^\dag{\mathbf{H}} \mathbf{c}^{(n)}\right) \delta\tau \\
&\quad + \mu N_\mathrm{w}^{(n)}\sqrt{\delta\tau} \check r^{(n)},
\end{align}
where the noise term averages to zero:
\begin{align}
\langle  \mu N_\mathrm{w}^{(n)}\sqrt{\delta\tau} \check r^{(n)} \rangle &= 0 . \label{eq:Nw_noise}
\end{align}
The noise term is written with the explicit factor $\sqrt{\delta\tau}$ to account for the fact that all individual steps in the random sampling procedure have a variance linear in $\delta\tau$. This is obtained from detailed inspection of the FCIQMC sampling process in Sec.~\ref{sec:noiseinfciqmc}, and is due to the fact that the individual random variables follow a scaled Bernoulli distribution and thus the variances are proportional to the mean. As the scalar noise in Eq.~\eqref{eq:scalarNorm} sums over many of such Bernoulli random variables sampled in each time step, we may assume that it is normally distributed as a consequence of the theorem of large numbers. We thus take $\check{r}^{(n)}$ as a normally distributed random variable with zero mean $\langle \check{r}^{(n)} \rangle =0$ and unit variance  $\langle (\check{r}^{(n)})^2 \rangle =1$. This is significant because the product $\Delta \check{W}^{(n)} \equiv \sqrt{\delta\tau}\, \check{r}^{(n)}$ has all the properties of a
Wiener increment \cite{Gardiner2009}. 
We have also included the factor $N_\mathrm{w}^{(n)}$ for later convenience and capture all remaining dependencies of the variance with the parameter $\mu$.

As long as we are in or close to the steady-state limit considered in Sec.~\ref{sec:exactpcb} we can assume that the quantity  $\tilde{\mathbf{1}}^\dagger\check{\mathbf{H}} \mathbf{c}^{(n)}$
can be written in terms of the norm projected energy $\bar{E}_{\tilde{\mathbf{1}}}$ of Eq.~\eqref{eq:E1tilde} as $\bar{E}_{\tilde{\mathbf{1}}} N_\mathrm{w}^{(n)}$.
We thus arrive at the stochastic difference equation formulated in terms of the walker number $N_\mathrm{w}^{(n)}$ and the shift $S^{(n)}$
\begin{align} \label{eq:Nwdiff}
N_\mathrm{w}^{(n+1)} &= N_\mathrm{w}^{(n)} + \left(S^{(n)} - \bar{E}_{\tilde{\mathbf{1}}} \right) N_\mathrm{w}^{(n)}  \delta\tau + \mu N_\mathrm{w}^{(n)} \Delta \check{W}^{(n)} .
\end{align}
Together with the deterministic update equation \eqref{eq:Supdate} for the shift, this provides a fully self-consistent model for the dynamics of the shift and walker number where the details of the random sampling process are compressed into the single parameter $\mu$. Assuming that $\mu$ is a constant parameter is a useful simplification that makes the model solvable. This assumption will be relaxed in App.~\ref{sec:normSDE}.



\subsection{Stochastic differential equation limit} \label{sec:SDElimit}


It is most convenient to convert the stochastic difference equation into a stochastic differential equation, which can then be treated with the powerful methods of stochastic calculus \cite{Gardiner2009}. Since  $\Delta \check{W}^{(n)}$ satisfies the properties of a Wiener increment, we can associate it with a well defined underlying Wiener process in a continuous time $t$ that is discretized into time steps of length $\delta\tau$. Then the coupled stochastic difference equations \eqref{eq:Nwdiff} and \eqref{eq:Supdate} can be identified as the
Euler-Maruyama 
discretization of the It\^{o} stochastic differential equation
\begin{align} \label{eq:dcscalar}
dN_\mathrm{w} &= \left[S(t)- \bar{E}_{\tilde{\mathbf{1}}} \right]N_\mathrm{w}(t)\, dt -  \mu
N_\mathrm{w}(t) \, d\check{W}(t) , \\
dS &= -\frac{\zeta}{\delta\tau} d \ln{N_w(t)} -\frac{\xi}{\delta\tau^2} \ln\frac{N_w(t)}{N_\mathrm{t}}\, dt
\end{align}
where $S(t)$ and $N_\mathrm{w}(t)$ are now continuous-time functions and $d\check{W}(t)$ is an infinitesimal Wiener increment. Note that this noise source satisfies the requirements of  It\^{o} calculus to be in the ``future'' of the dynamical variables $N_\mathrm{w}(t)$ and $S(t)$, because the noise in the FCIQMC sampling procedure that generated it [also see Eq.~\eqref{eq:scalarNorm}] is generated by fresh random numbers in each time step that do not depend on the instantaneous values of ${N_\mathrm{w}^{(n)}}$ or $S^{(n)}$.

It is now convenient to introduce a variable transformation and replace the walker number with the new variable
\begin{align}
    x(t) = \ln\frac{N_\mathrm{w}(t)}{N_\mathrm{t}}.
\end{align}
In the first place this transformation is convenient because it removes the logarithm terms in the shift update equation. However, it also serves a second more important purpose in removing the product of fluctuating variable on the right hand side of Eq.~\eqref{eq:dcscalar}, as we will see.

In It\^o calculus one has to be careful when performing variable transformations and counting orders of differentials. This is to account for the fact that the standard deviation of the Wiener increment gives $\sqrt{\delta\tau}$. The resulting procedure is known as  It\^o's Lemma \cite{enwiki:1008274165}. The It\^o rules for the infinitesimal increments are
\begin{align} \label{eq:ItoRules}
d\check{W}^2 &= dt ,\\
dt^2 = d\check{W}\,dt &= 0 .
\end{align}
Recall that
\begin{align}
df(y) = f'(y)\,dy + \frac{1}{2} f''(y)\, dy^2 + \ldots ,
\end{align}
and thus
\begin{align} \label{eq:changeofvars}
dx & = \frac{1}{N_\mathrm{w}} dN_\mathrm{w} - \frac{1}{2 N_\mathrm{w}^2} dN_\mathrm{w}^2  + \ldots .
\end{align}
Performing the variable transformation by inserting Eq.\ \eqref{eq:dcscalar} for $dN_\mathrm{w}$ into Eq.\ \eqref{eq:changeofvars} and applying It\^o rules yields
\begin{align} \label{eq:scalarSDE1}
 dx &= \left[S(t)- \tilde E -\frac{1}{2}  \mu^2\right]\, dt -\mu \, d\check{W}(t) , \\ \label{eq:scalarSDE2}
dS &= - \frac{\zeta}{\delta\tau}\, dx - \frac{\xi}{\delta\tau^2}  x(t)\, dt ,
\end{align}
which is the final form of the coupled It\^{o} stochastic differential equations (SDEs).


In the limit where no noise is present, $\mu =0$, the above SDE simplifies to a set of coupled linear ordinary differential equations (ODEs)
\begin{align}
\frac{dx}{dt} &= S(t) - \tilde{E} , \\
\frac{dS}{dt} & = - \frac{\zeta}{\delta\tau}\, \frac{dx}{dt} - \frac{\xi}{\delta\tau^2}  x(t) .
\end{align}
These equations were previously derived for the population dynamics of FCIQMC in Ref.~\cite{Yang2020},
and
describe the motion of a damped harmonic oscillator for $x(t)$. The equilibrium solution (and global attractor) is $x(t)=0$ and $S(t) = \tilde{E}$. Note that $x=0$ means $N_\mathrm{w} = N_\mathrm{t}$, i.e.~the walker number reaches the pre-set target walker number.
The two fundamental solutions of the ODE are exponentials $x_\pm(t) = \exp(t/T_\pm)$ with time constants $T_\pm = \delta\tau(\zeta \pm \sqrt{\zeta^2-4\xi})/(2\xi)$.

\subsection{Population control bias for the shift estimator}\label{sec:steadystateSDE}
In the previous section we have seen that without noise the time evolution is given by an exponential decay to a steady state. In the presence of noise the long-time limit will not be time-independent but sees fluctuations of $x(t)$ and $S(t)$ around some mean values. Taking the ensemble average in this steady-state situation, we can thus expect to have
\begin{align}
\langle dx \rangle &= 0 ,\\
\langle dS \rangle &= 0 .
\end{align}
The Wiener increment by definition fulfills $\langle dW \rangle = 0$. Thus, taking the average of the coupled SDEs \eqref{eq:scalarSDE1} and \eqref{eq:scalarSDE2}, we obtain
\begin{align}\label{eq:scalarPCB1}
\langle S \rangle -\tilde{E} &= \frac{1}{2} \mu^2 ,\\ \label{eq:scalarPCB2}
\langle x \rangle &= 0 .
\end{align}
The first equation yields an explicit expression for the population control bias in the scalar model, i.e. the deviation of the averaged shift $\langle S \rangle$ from the target energy $\tilde{E}$. The second equation asserts that the
walker number $N_\mathrm{w}$ fluctuates around the target walker number $N_\mathrm{t}$.

\subsection{Steady-state solution of the SDE}

Because the coupled SDEs \eqref{eq:scalarSDE1} and \eqref{eq:scalarSDE2}  are linear, their general solutions can be found with a Greens function technique as shown in App.~\ref{sec:GreensFunctions}. Here we are specifically interested in the
long time limit where the only dynamics left is due to the injected noise. In this case the solutions for the logarithmic walker number and the shift can be written as
\begin{align}\label{eq:x_sol}
x(t) &= -\mu \int_{-\infty}^{t} g_{11}(t-t') dW(t') , \\ \label{eq:s_sol}
S(t) &= \tilde{E} +\frac{1}{2}\mu^2 - \mu \int_{-\infty}^{t} g_{21}(t-t') dW(t') .
\end{align}
The Greens functions for the case of critical damping, where $4\xi = \zeta$, read
\begin{align} \label{eq:g11_crit}
g_{11}(t) & = \theta(t) (1-\gamma t) e^{-\gamma t}, \\  \label{eq:g21_crit}
g_{21}(t) & =  \theta(t) (-2\gamma + \gamma^2 t) e^{-\gamma t} ,
\end{align}
where $\gamma = \zeta/(2\delta\tau)$ is the damping constant. More general expressions for the overdamped and underdamped case are derived in App.~\ref{sec:GreensFunctions}.
Note that direct averaging of Eqs.~\eqref{eq:x_sol} and \eqref{eq:s_sol}  yields the correct averages \eqref{eq:scalarPCB1} and \eqref{eq:scalarPCB2} that were previously obtained directly from the SDE.

\subsection{Covariances and correlation functions in the scalar model} \label{sec:correlations}

The full solutions allow us to go further, though, and
derive any correlation function. Starting with auto-covariance functions with a time lag of $h$, we obtain for the
logarithmic walker number $x$
\begin{align} \label{eq:x-x}
\operatorname{cov}[ x(t-h), x(t)] &= \frac{\mu^2}{4} \left(\frac{1}{\gamma} - |h|\right) e^{-\gamma |h|} ,
\end{align}
and for the shift $S$
\begin{align} \label{eq:av-s}
\operatorname{cov}[S(t-h), S(t)]= \frac{\mu^2}{4} (5\gamma -3\gamma^2 |h|)  e^{-\gamma |h|} .
\end{align}
For the cross-covariances we obtain (for $h\ge 0$):
\begin{align} \label{eq:cv-xs}
\operatorname{cov}[x(t-h), S(t) ] &= -\frac{\mu^2}{4}(2 - \gamma h)  e^{-\gamma h}, \\
\label{eq:cv-xxs}
\operatorname{cov}[x(t), S(t-h)] &= -\frac{\mu^2}{4}(2 - 3\gamma h)  e^{-\gamma h}.
\end{align}
Specifically, for the cross-covariance without lag ($h=0$) we recover the value of the population control bias in the shift:
\begin{align} \label{eq:covxS}
\operatorname{cov}(x,S) = \frac{1}{2}\mu^2 = \langle S \rangle - \tilde{E} .
\end{align}

Covariances of the walker number ${N_\mathrm{w}(t)} = {N_\mathrm{t}} \exp[x(t)]$ can easily be obtained
when the fluctuations of the walker number are small, i.e.\ $\mathrm{std}({N_\mathrm{w}})\ll {N_\mathrm{w}}$ from the method of small increments. For arbitrary $A(t)$, e.g., we find
\begin{align}
\operatorname{cov}[N_\mathrm{w}(t), A(t)] \approx  N_\mathrm{t} \operatorname{cov}[x(t), A(t)],
\end{align}
in this regime, where also $\langle N_\mathrm{w}\rangle = N_\mathrm{t}$. Specifically, this provides us with an expression for the equal-time covariance of the walker number and the shift [from Eq.\ \eqref{eq:covxS}]
\begin{align}
\operatorname{cov}(N_\mathrm{w},S) = \frac{1}{2}N_\mathrm{t} \mu^2 .
\end{align}
With this information, we can thus determine the norm-projected energy estimator for the scalar model from Eq.\ \eqref{eq:nEnergy} as
\begin{align}\label{eq:npe_sm}
\bar{E}_{\tilde{\mathbf{1}}} = \left\langle S \right\rangle - \frac{1}{2}\mu^2  = \tilde{E}.
\end{align}
I.e.\ the value is equal to the exact energy. This is also consistent with directly interpreting the definition of the norm projected energy of Eq.\ \eqref{eq:E1tilde} in the context of the scalar model, where it necessarily reproduces the exact energy $\tilde{E}$.
We can thus conclude that the norm-projected energy estimator is not affected by the population control bias in the scalar model.

\subsection{Evaluation of the scalar model and comparison to full FCIQMC}

The scalar model in the form of the SDEs \eqref{eq:dcscalar} turned out to be exactly solvable, which provides a convenient source of insight into the population control bias in the shift estimator and its parameter dependences. The population control bias for the shift estimator was found to be completely determined by the coefficient $\mu$ describing the strength of the noise source by Eq.\ \eqref{eq:scalarPCB1}. In particular, we find no  dependence on the population control parameters $\zeta$ (for damping) or $\xi$ (for forcing) that appear in Eq.\ \eqref{eq:Supdate}. Also, since the time step drops out of the differential equation model, there is no dependence on $\delta\tau$. All of this is consistent with empirical observations on numerical FCIQMC simulation, as detailed in Appendix \ref{sec:params} (see Figs.\ \ref{fig:estimatorswithxi} and \ref{fig:biasdeltatau}).

Unfortunately, the scalar model is not useful for studying the population control bias in the projected energy estimators, as the only available projected energy estimator is unbiased according to Eq.\ \eqref{eq:npe_sm}.

\subsubsection{Nonlinear extension of the scalar model}
The scalar model of Eqs.\ \eqref{eq:dcscalar} also makes no predictions for the dependence of the population control bias on the walker number $N_\mathrm{t}$. A more careful analysis of the FCIQMC sampling procedure is necessary to obtain this information. In Sec.\ \ref{sec:noiseinfciqmc} this will be done taking into account the structure of specific Hamiltonian matrices. Within the scalar model
we can obtain a more realistic description by replacing
\begin{align} \label{eq:nonlinearnoise}
\mu N_\mathrm{w}  \to\eta \sqrt{N_\mathrm{w}} ,
\end{align}
in Eq.\ \eqref{eq:dcscalar}, with some constant $\eta$.
This can be motivated by the assumption that the collective action of the sampling procedure can be treated as a sum of Bernoulli random variables, for which the variance is proportional to the mean. While the resulting, modified, scalar model is nonlinear and more complicated than the previous version, the steady state averages can be obtained in analogy to Sec.\ \ref{sec:steadystateSDE} and yield
\begin{align}
\langle S \rangle -\tilde{E} &= \frac{\eta^2}{2 N_\mathrm{t}}.
\end{align}
The prediction of this nonlinear scalar model is thus that the population control bias in the shift estimator scales with $N_\mathrm{t}^{-1}$.
This has been seen in projector Monte Carlo calculations in many cases before, although the evidence presented in Sec.\ \ref{sec:evidence} gave examples to the contrary. In the more detailed analysis of Sec.\ \ref{sec:noiseinfciqmc} we will see that the assumption \eqref{eq:nonlinearnoise} is too simplistic, and indeed the details of the Hamiltonian matrix have to be taken into account.

\subsubsection{Comparing correlation functions from FCIQMC and the scalar model}
\begin{figure}
\begin{tabular}{c}
(a) \\
\includegraphics[width=\columnwidth]{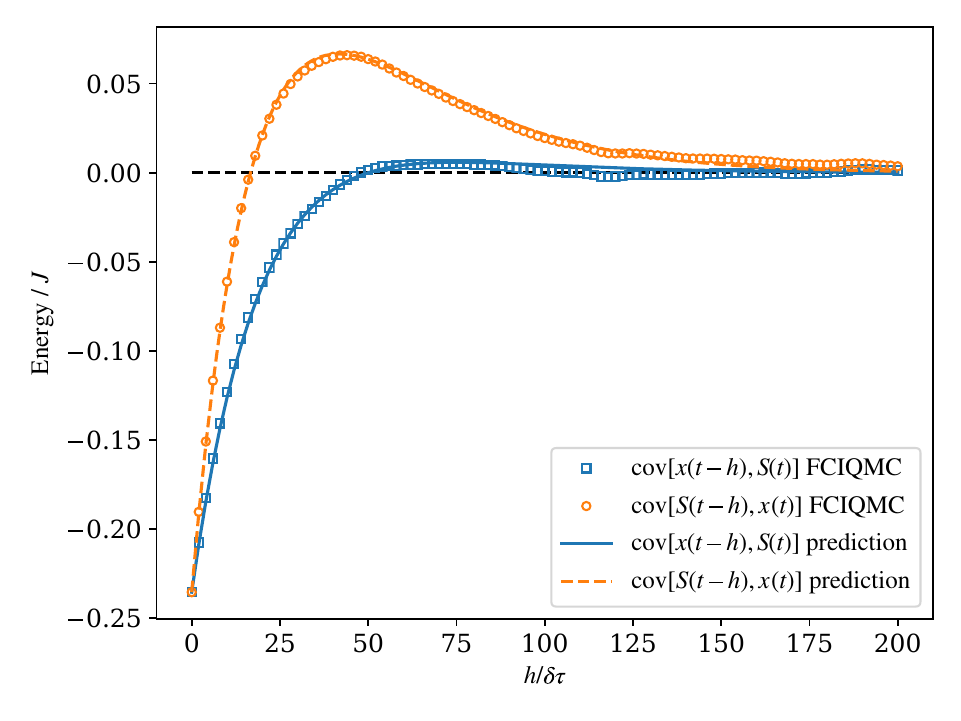}\\
(b)\\
\includegraphics[width=\columnwidth]{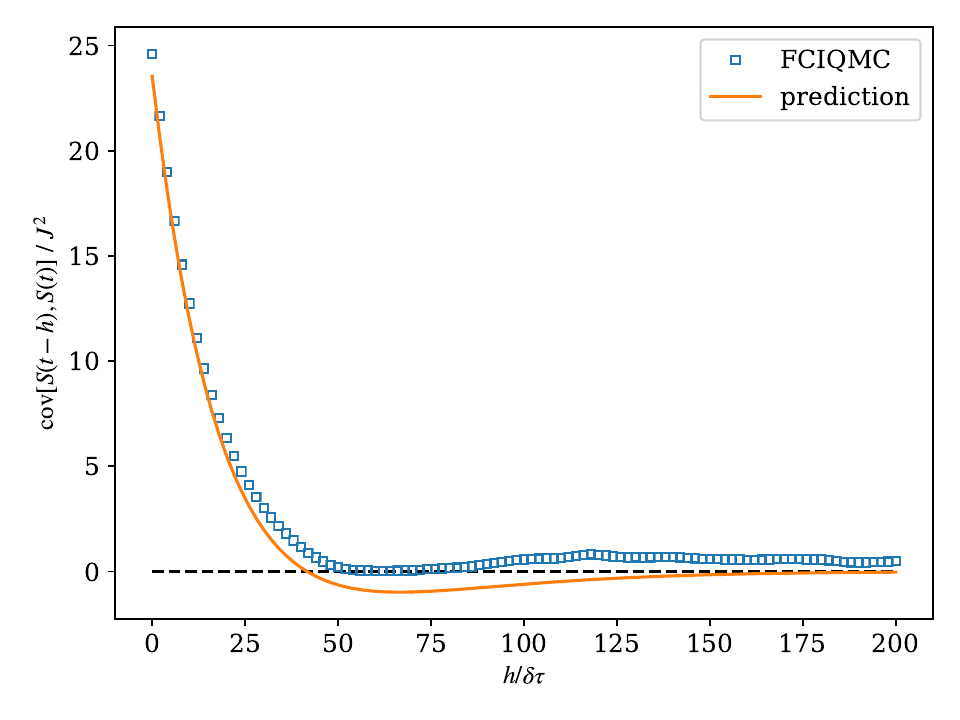}
\end{tabular}
\caption{Covariances vs.\ time delay in the well-sampled regime. Panel
(a) shows the cross-covariances $\operatorname{cov}[x(t-h), S(t) ] $ and $\operatorname{cov}[S(t-h), x(t)]$ and panel (b) shows the auto-covariance $\operatorname{cov}[S(t-h), S(t)]$.
%
%
Symbols depict numerical (``FCIQMC'') results with $N_\mathrm{t} = 100$ walkers for a Bose-Hubbard chain with $N=M=10$, $U/J=6$.
These numerical results are fairly well matched by
the analytical results (``predictions'') of Eqs.~\eqref{eq:cv-xs} in panel (a), and Eq.~\eqref{eq:av-s} in panel (b) from the scalar model shown as lines.
The decay rate in the scalar model $\gamma = \zeta/2$ is already determined by  the value $\zeta = 0.08$ used in the numerical calculation for population control and is not a free parameter. It fully determines the correlation time scale  $\delta\tau/\gamma = 2\delta\tau/\zeta = 25 \delta\tau$.
The remaining parameter $\mu$ was set to $\mu^2/2 = -\operatorname{cov}(S,x)$ as obtained from the numerical results.
Averaging was performed over $\Omega = 2^{18}$ time steps with $\delta\tau = 0.001 J^{-1}$ after equilibrating for 5,000 time steps.  The forcing parameter was set to $\xi = \zeta^2/4$.} \label{fig:cross-cov}
\end{figure}

The exactly solvable scalar model makes interesting predictions for correlation functions of the time series with Eqs.\ \eqref{eq:x-x} to \eqref{eq:cv-xxs}.
The analytical results for the  cross-covariance  and the auto-covariance of the shift are compared to numerical results of an FCIQMC calculation with $N_\mathrm{t} = 100$ in a Hilbert space of $\approx 9\times10^4$  in Fig.~\ref{fig:cross-cov}.
It is striking to see the predictions derived from the scalar model capturing the behavior of the cross-covariance functions in panel (a) almost perfectly. The only free model parameter $\mu$ was adjusted to the value of the equal-time covariance. The agreement is less perfect for the autocovariance of the shift in Fig.~\ref{fig:cross-cov}(b), where the scalar model predicts a zero crossing at $h=5/(3\gamma)$, which is not seen in the FCIQMC data.

\begin{figure}
\begin{tabular}{c}
(a) \\
\includegraphics[width=\columnwidth]{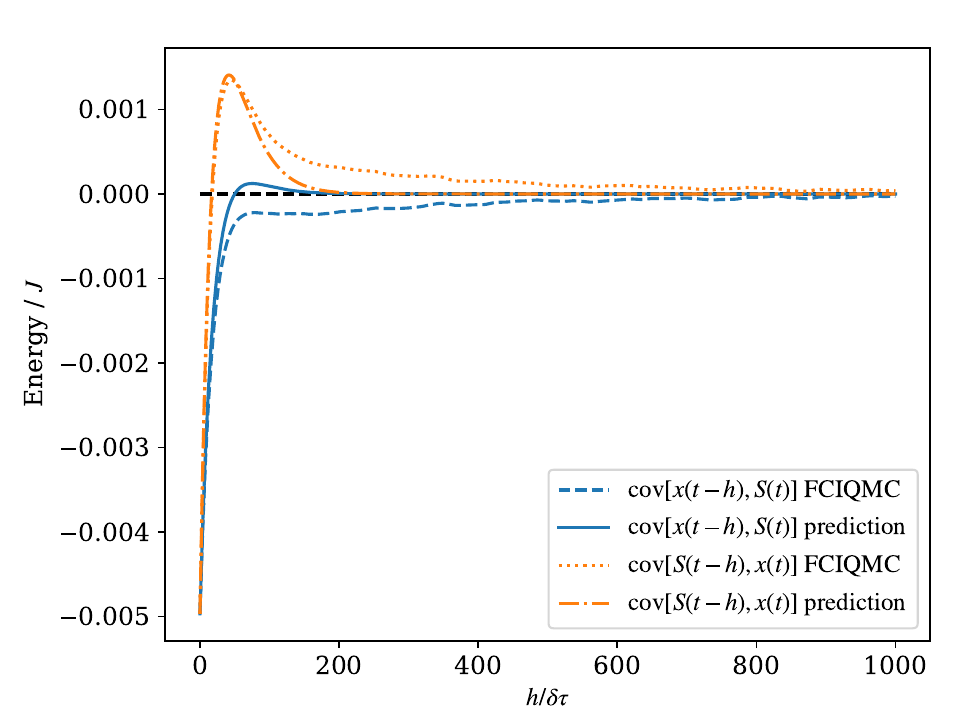}\\
(b)\\
\includegraphics[width=\columnwidth]{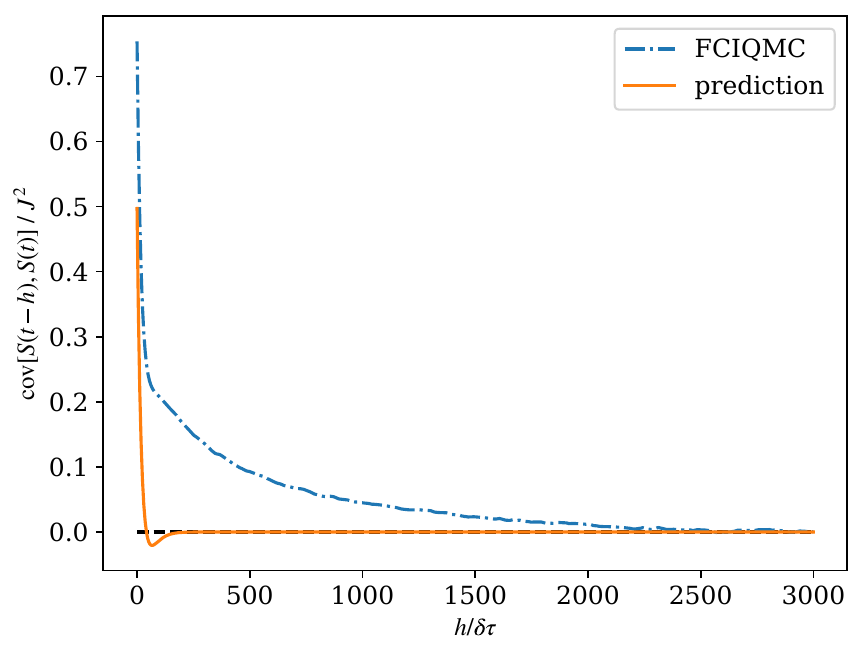}
\end{tabular}
 \caption{Evidence for long correlation times. Cross-covariances $\operatorname{cov}[x(t-h), S(t)]$ and $\operatorname{cov}[S(t-h), x(t)]$ and the auto-covariance of the shift $\operatorname{cov}[S(t-h), S(t)]$ as function of the time delay are shown in panels (a) and (b), respectively. Numerical data (``FCIQMC'') and analytical predictions from Eqs.\ \eqref{eq:cv-xs} and \eqref{eq:av-s} (``prediction'') are shown as per legends.
The FCIQMC results for the real-space Bose-Hubbard chain with $N=M=20$, $U/J=6$ with $N_\mathrm{t} = 10,000$ walkers demonstrate a much longer time scale for the decay of the correlation functions than the analytical prediction of $\delta\tau/\gamma = 2\delta\tau/\zeta = 25 \delta\tau$. The value $\zeta = 0.08$ at critical forcing $\xi = \zeta^2/4$ was used for population control in FCIQMC.
The parameter $\mu$ was set to $\mu^2/2 = -\operatorname{cov}(S,x)$ as obtained from the numerical cross-correlation at delay $h=0$.
 %
%
 Averaging was performed over  $\Omega=4\times10^6$ time steps with $\delta\tau = 0.001 J^{-1}$ after equilibrating for 50,000 time steps. Statistical error bars for the numerical results (not shown) are comparable to the line width.
 } \label{fig:cov2}
\end{figure}

The situation changes dramatically in a system with a larger Hilbert space as can be seen in Fig.\ \ref{fig:cov2}. Here the dimension of Hilbert space is $\binom{M+N-1}{N} \approx 7 \times 10^{10}$, almost seven orders of magnitude larger than the walker number  $N_\mathrm{t} = 10,000$. While the correlation functions now show a fast initial decay consistent with the analytical prediction (here  $\delta\tau/\gamma = 2\delta\tau/\zeta = 25 \delta\tau$), the eventual decay of the correlation functions to zero is dominated by a second, much longer time scale of the order of $10^3 \delta\tau$. We attribute this longer time scale to the time it takes for the walker population to explore Hilbert space. Eventually, this correlation time is bounded by the Poincar\'e recurrence time  of the sampling process, which can become very large in a large Hilbert space.
Clearly, such effects are not captured in the scalar model and its predictions of Eqs.\ \eqref{eq:x-x} to \eqref{eq:cv-xxs}, because the noise was modeled by an uncorrelated scalar source term [see Eq.\ \eqref{eq:Nwdiff}].  The scalar model could thus be made more realistic by injecting  noise with a finite correlation time.

\section{Unbiased estimators by reweighting} \label{sec:Reweighting}

In section \ref{sec:exactpcb} we argued that the population control bias originates from the non-vanishing covariance of the fluctuating shift and coefficient vector in the product of the two quantities in the FCIQMC master equation \eqref{eq:Cupdate} [see Eq.\ \eqref{eq:Sc_cov}]. Hetherington \cite{Hetherington1984} first discussed the construction of unbiased estimators from weighted averages, which was put into practice in Refs.\ \cite{Nightingale1986,Nightingale1988}. We briefly review the construction of unbiased estimators following Ref.\ \cite{Umrigar1993} (see also \cite{Ghanem2021}).

\subsection{The reweighting procedure} \label{sec:Reweighting_A}

Suppose we replace the update equation for the coefficient vector \eqref{eq:Cupdate} by
\begin{align}
\label{eq:fupdate}
\mathbf{f}^{(n+1)} & = [\mathds{1} + \delta\tau(E_\mathrm{f}\mathds{1} - \check{\mathbf{H}})]\mathbf{f}^{(n)} ,
\end{align}
where the fluctuating shift $S^{(n)}$ is replaced by a constant $E_\mathrm{f}$. This equation is not practical for forward propagation because it is unstable to exponential growth or decay. Let us assume, however, that $E_\mathrm{f}$ is
chosen such that the norm $\lVert\mathbf{f}^{(n_\mathrm{f})}\rVert_1$ takes a given desired value at the final point $n_\mathrm{f}$ of a particular time series. Then the covariance $\mathrm{cov}(E_\mathrm{f},\mathbf{f}^{(n)})$ vanishes trivially. Thus there is no population control bias, and the expected value $\langle\mathbf{f}\rangle$ becomes collinear to the exact ground state coefficient vector for sufficiently large $n_\mathrm{f}$.

The idea of the reweighting procedure is to approximately generate the time series for the vectors $\mathbf{f}^{(n)}$ (or derived quantities). This is achieved by undoing the effect of the fluctuating shift $S^{(n)}$ on a given time series of $\mathbf{c}^{(n)}$ and $N_\mathrm{w}^{(n)}$, obtained using the standard procedure of Eqs.\ \eqref{eq:Cupdate} and \eqref{eq:Supdate} .

The effect of the shift can be undone for a single time step using 
\begin{align}
\nonumber
\mathds{1} + \delta\tau\left(E_\mathrm{f}\mathds{1} - \check{\mathbf{H}}\right) &= \exp\left[\delta\tau \left(E_\mathrm{f} - S^{(n)}\right)\right] \times\\
&\quad  \left[\mathds{1} + \delta\tau\left(S^{(n)}\mathds{1} - \check{\mathbf{H}}\right)\right]
+ \mathcal{O}\left(\delta\tau^2\right) ,
\end{align}
where we have used the expansion of the exponential function $\exp(x) = 1+x+\mathcal{O}(x^2)$.
The effect of ${\tilde{h}}$ steps can consequently be undone by multiplying with a weighting factor made up of products of exponential factors up to a small error of order $\delta\tau^2$.
Defining the weight factor
\begin{align}
w_{\tilde{h}}^{(n)} = \prod_{j=1}^{{\tilde{h}}} \exp\left[\delta\tau\left(E_\mathrm{f} - S^{(n-j)}\right)\right] ,
\end{align}
the quantity
\begin{align}\label{eq:f_weighted}
\mathbf{f}^{(n)} & = w_{n-n_0}^{(n)} \mathbf{c}^{(n)} ,
\end{align}
approximately fulfills the iteration equation \eqref{eq:fupdate} for $n\ge n_0$  with initial condition $\mathbf{f}^{(n_0)}=\mathbf{c}^{(n_0)}$ and is unbiased for $n-n_0\to\infty$. Unbiased estimators for observables can thus be obtained by simply replacing the coefficient vector $\mathbf{c}^{(n)}$ by $w_{\tilde{h}}^{(n)}\mathbf{c}^{(n)}$ in the corresponding expressions with a suitably chosen reweighting depth ${\tilde{h}}$. 

An asymptotically unbiased estimator for the ground state energy based on the projected energy of Eq.\ \eqref{eq:p_energy}, termed ``mixed estimator'' in Ref.\ \cite{Umrigar1993}, can be defined as
\begin{align} \label{eq:Emix}
E_\mathrm{mix}({\tilde{h}}) = \frac{\sum_n w_{\tilde{h}}^{(n)} \mathbf{y}^\dag \mathbf{H} \mathbf{c}^{(n)} }{\sum_n w_{\tilde{h}}^{(n)} \mathbf{y}^\dag \mathbf{c}^{(n)} }
\end{align}
where the sum runs over a sufficiently large part of an equilibrated time series (the sample).
Note that for ${\tilde{h}}=0$ no reweighting takes place and instead we recover the
projected energy of Eq.\ \eqref{eq:p_energy}: $\langle E_\mathrm{mix}(0)\rangle = \bar{E}_\mathbf{y}$.
The expected value of the estimator $E_\mathrm{mix}({\tilde{h}})$ is unbiased in the limit ${\tilde{h}}\to\infty$ and $\delta\tau\to 0$.
However, the variance grows with ${\tilde{h}}$, which makes it impractical to take the large ${\tilde{h}}$ limit in numerical calculations.

Note that the actual value of the constant $E_\mathrm{f}$ is not very important as it can easily be seen that the value $E_\mathrm{mix}({\tilde{h}})$ is independent of $E_\mathrm{f}$. In order to minimize rounding errors it should be chosen such as to avoid extremely large or small weight factors and thus we set $E_\mathrm{f}$ to the sample mean of the shift.

The time series for the walker number $N_\mathrm{w}^{(n)}$ can likewise be unbiased by reweighting with the weight factors $w_{\tilde{h}}^{(n)}$ [following from Eq.\ \eqref{eq:f_weighted}]. An asymptotically unbiased version of the growth estimator from Eq.\ \eqref{eq:growth_witness_biased} is given by \cite{Umrigar1993}
\begin{align} \label{eq:Egr}
E_\mathrm{gr}({\tilde{h}}) = E_\mathrm{f} - \frac{1}{\delta\tau}\ln\frac{\sum_n w_{\tilde{h}+1}^{(n+1)} N_\mathrm{w}^{(n+1)}}{\sum_n w_{\tilde{h}}^{(n)} N_\mathrm{w}^{(n)}} .
\end{align}
It is easy to show that the growth estimator without reweighting (${\tilde{h}}=0$) is approximately equal to a time series average over $G^{(n)}$ of Eq.\ \eqref{eq:growth_witness_biased}:  $\langle E_\mathrm{gr}(0)\rangle = \langle G\rangle +\mathcal{O}(\delta\tau^2)$, which is obtained in the limit of summing over a long time series. By the arguments of Sec.~\ref{sec:normprojected}, the growth estimator 
thus becomes equivalent to the norm-projected energy estimator $\bar{E}_{\tilde{\mathbf{1}}} $ of Eq.\ \eqref{eq:E1tilde}.

The growth estimator is closely related to the shift estimator (sample mean of $S^{(n)}$) and can be understood as the improved and reweighted version of the shift. Since the weight factors asymptotically remove the bias from the time series of the walker numbers, $E_\mathrm{gr}({\tilde{h}})$ is formally unbiased in the limit ${\tilde{h}}\to\infty$ and $\delta\tau\to 0$.

\subsection{Reweighting for the scalar model}

The effect of the reweighting procedure cannot be meaningfully studied in the scalar model because the relevant energy estimators are already unbiased for ${\tilde{h}}=0$, i.e.\ without actual reweighting.

\subsection{Analysis of the unbiased estimators}

\begin{figure}
\includegraphics[width=\columnwidth]{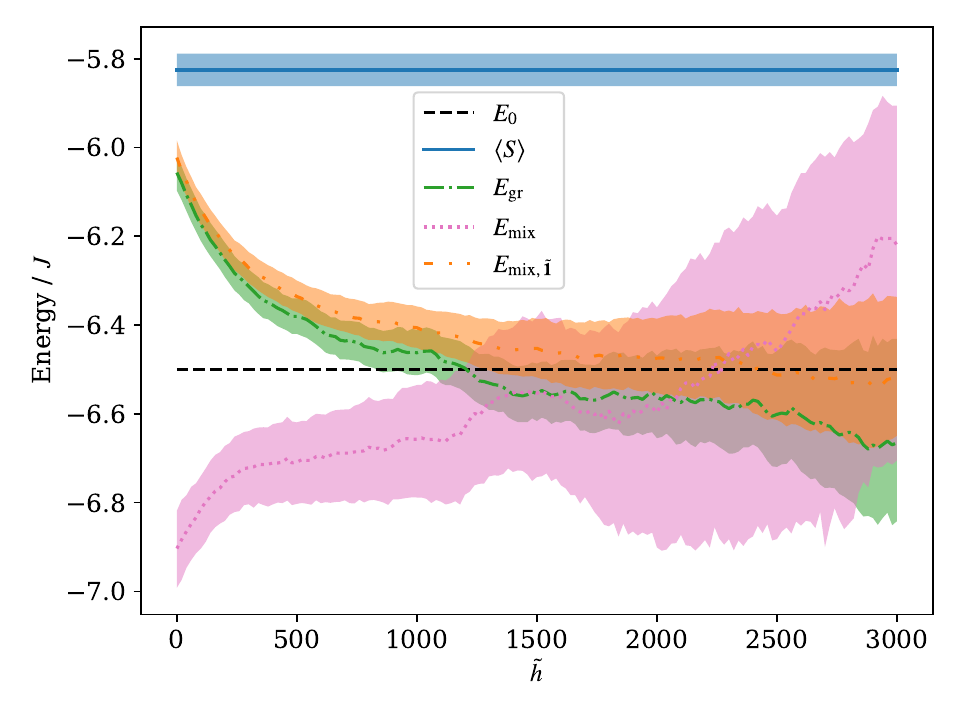}
 \caption{\rev{Reweighted estimators for  $N=10$, $M=10$ Bose Hubbard model and  $N_\mathrm{t} = 100$ walkers. The growth estimator $E_\mathrm{gr}$ of Eq.~\eqref{eq:Egr} (green dash-dotted line) and the mixed estimator of Eq.~\eqref{eq:Emix} from a single configuration trial vector ($E_\mathrm{mix}$, dotted magenta line) and from the norm projector ($E_{\mathrm{mix},\tilde{\mathbf{1}}}$, dash-dot-dotted orange line) are shown as a function of the reweighting depth ${\tilde{h}}$ in comparison to the shift estimator $\langle S\rangle$ (blue solid line), and the exact energy $E_0$ (dashed black line).
The time scale ${\tilde{h}}\, \delta\tau$ can be directly compared to the delay $h$ of the correlation functions shown in Fig.~\ref{fig:cross-cov} and to the decorrelation time scale $\sim 2^{11}\delta\tau=2048\,\delta\tau$ estimated from reblocking of the shift time series with $\Omega=2^{20} \approx 10^6$ steps. All other parameters as in Fig.~\ref{fig:cross-cov}.}
 } \label{fig:rew10}
\end{figure}

\begin{figure}
\includegraphics[width=\columnwidth]{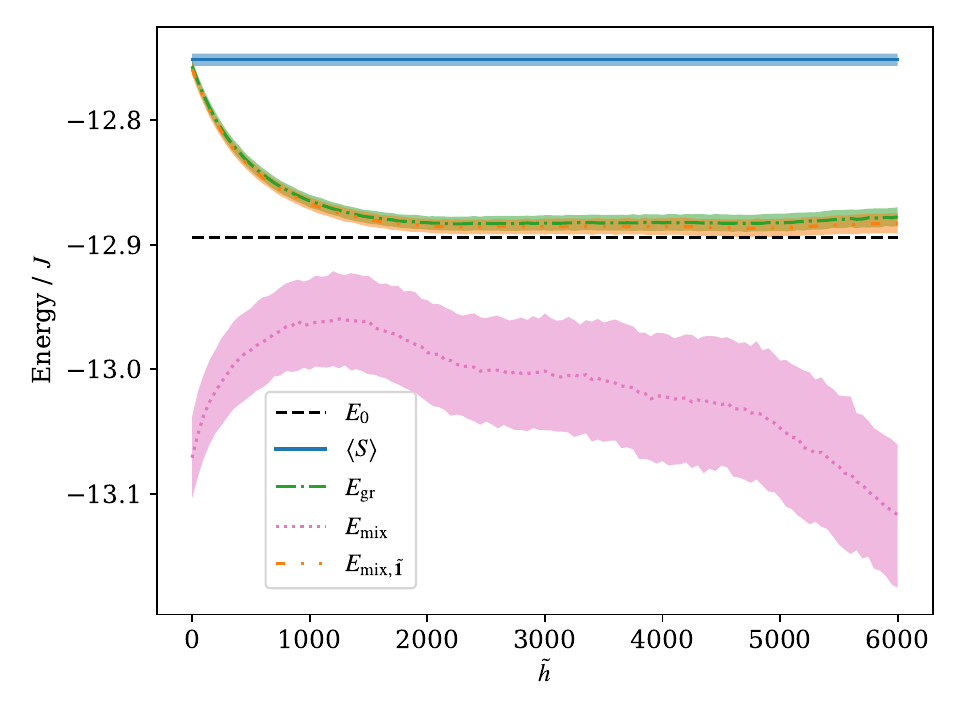}
 \caption{\rev{Reweighted estimators for $N=20$, $M=20$ Bose-Hubbard model and $N_\mathrm{t} = 10^4$ walkers (c.f.\ Figs.~\ref {fig:variationalenergy} and \ref{fig:cov2}).
The growth estimator $E_\mathrm{gr}$ of Eq.~\eqref{eq:Egr} (green dash-dotted line) and the mixed estimator of Eq.~\eqref{eq:Emix} from a single configuration trial vector ($E_\mathrm{mix}$, dotted magenta line)  and from the norm projector ($E_{\mathrm{mix},\tilde{\mathbf{1}}}$, dash-dot-dotted orange line) are shown as a function of the reweighting depth ${\tilde{h}}$ in comparison to the shift estimator $\langle S\rangle$ (blue solid line), and the (extrapolated) exact energy $E_0$ (dashed black line). The time scale ${\tilde{h}}\, \delta\tau$ can be directly compared to the delay $h$ of the correlation functions shown in Fig.~\ref{fig:cov2} and to the decorrelation time scale $\sim 2^{12}\delta\tau=4096\,\delta\tau$ estimated from reblocking of the shift time series with $\Omega=10^7$ steps. All other parameters as in Fig.~\ref{fig:cov2}.}
 } \label{fig:rew20}
\end{figure}

Figures \ref{fig:rew10} and \ref{fig:rew20} show the reweighted estimators for FCIQMC calculations in Hilbert spaces of different size.
The data shows how initially the reweighted estimators move closer to the exact result as the  reweighting depth ${\tilde{h}}$ is increased. 
\rev{Both the growth estimator $E_\mathrm{gr}$ and the mixed energy estimator relating to the norm-projected energy $E_{\mathrm{mix},\tilde{\mathbf{1}}}$ lie above the exact energy for $\tilde{h}=0$ and initially show a downward trend towards the exact energy when increasing $\tilde{h}$, as expected. The same behavior was also observed in Ref.~\cite{Umrigar1993}, where an argument is made that the mixed estimator in diffusion Monte Carlo (equivalent to $E_{\mathrm{mix},\tilde{\mathbf{1}}}$) always has a positive population control bias, i.e. provides an upper bound to the exact energy. For the mixed estimator from a single configuration trial vector $E_\mathrm{mix}$ we see different behavior, as it lies below the exact energy.
Generally, we see non-monotonic behavior for larger ${\tilde{h}}$ and growing error bars estimated from blocking analysis. This is consistent with
}
%
what has been reported previously in the literature, e.g.\ in Ref.~\cite{Vigor2015}.

\rev{It can be seen in Figs.\ \ref{fig:rew10} and \ref{fig:rew20} that the growth estimator $E_\mathrm{gr}$ and the mixed energy estimator relating to the norm-projected energy $E_{\mathrm{mix},\tilde{\mathbf{1}}}$ show very similar results, agreeing within error bars. This extends the finding of Sec.\ \ref{sec:Reweighting_A} that both are approximately equivalent. 
Note that the growth estimator is much easier to compute as only the time series data of the shift and walker number are needed, which are calculated anyway. In contrast, computing the norm-projected energy through the local energy is more costly because it requires an additional matrix-vector multiplication $\mathbf{H} \mathbf{c}^{(n)}$ in each time step
\footnote{\rev{In diffusion Monte Carlo with importance sampling, the mixed estimator using the guiding function as the trial wave function (or projector in our language) is computed from the local energy (see e.g.~\cite{Umrigar1993}).
Because the local energy is calculated as part of the diffusion Monte Carlo algorithm, obtaining the data required for the mixed estimator comes at no extra cost. 
In the context of this work, where no importance sampling is used, the norm projector $\tilde{\mathbf{1}}$ plays the role of the guiding function and the relevant mixed estimator becomes $E_{\mathrm{mix},\tilde{\mathbf{1}}}$.
It is observed in practice that the mixed estimator in diffusion Monte Carlo is 
essentially equivalent to the growth estimator \cite{Umrigar:private}. This is consistent with the arguments in the Sec.~\ref{sec:normprojected}, which  straightforwardly  generalise to include importance sampling.}}.
The mixed energy estimator obtained from a single configuration is comparatively easier to compute but its quality as an energy estimator suffers from small overlaps in the sparsely sampled coefficient vectors (See also Ref.~\cite{Petruzielo2012} for a study of different quality projectors.).
}

It is interesting to think of ${\tilde{h}}\, \delta\tau$ as a time scale and compare Figs.~\ref{fig:rew10} and \ref{fig:rew20} to the correlation functions shown in the corresponding Figs.~\ref{fig:cross-cov} and \ref{fig:cov2}. For both cases, the time scale of correlations induced by the walker number control procedure is $\delta\tau/\gamma = 25 \delta\tau$ as per Sec.~\ref{sec:correlations}, which is much shorter than the time scale on which the reweighting is efficient. It rather appears that the longer time scale of $\sim 3,000$ time units observed for the decay of correlations in Fig.~\ref{fig:cov2} is relevant for the reweighted estimators, even though it is not evident in the correlation functions of the smaller system in Fig.~\ref{fig:cross-cov}. 
However, we also see non-monotonic behavior and significant growth of error bars on that time scale. 

Reference \cite{Nightingale1988} suggested to choose the time scale ${\tilde{h}}\, \delta\tau$ such that detected autocorrelations in the Monte Carlo time series have decayed below a statistically significant level. \rev{Estimating the decorrelation time from reblocking \cite{Flyvbjerg1989} combined with hypothesis testing \cite{Jonsson2018} yielded $\approx 2000$ steps for Fig.~\ref{fig:rew10} and $\approx 4000$ steps for Fig.~\ref{fig:rew20}. 
For the example with small walker number $N_\mathrm{t} = 100$ shown in  Fig.~\ref{fig:rew10} all estimators agree with the exact energy within error bars at the decorrelation time scale, but stochastic error bars showing 68\% confidence intervals are growing quite rapidly. For the larger walker numbers in Fig.~\ref{fig:rew20} the statistical errors grow more slowly. In particular the mixed estimator from projection onto a single configuration becomes non-monotonic and deteriorates before the decorrelation time scale.}
Accurate prediction of the optimal reweighting depth may thus require further study.

\section{Noise in the stochastic FCIQMC algorithm} \label{sec:noiseinfciqmc}

In this section we model the noise generated in the FCIQMC sampling process in the sparse walker regime on the level of individual matrix elements. This allows us to derive explicit relations for the shift and the projected energy estimators for specific cases of the Bose Hubbard Hamiltonian.

\subsection{FCIQMC sampling approximated by Wiener process}

The random processes associated with the individual steps of the matrix-vector multiplication in the walker update equation \eqref{eq:Cupdate} are considered in detail in  App.~\ref{sec:sampling} for the integer walker FCIQMC algorithm of Ref.~\cite{Booth2009}. This analysis suggest the following representation of the walker update
\begin{align} \label{eq:FCIQMCWiener}
\mathbf{c}^{(n+1)} -  \mathbf{c}^{(n)}  & = (S^{(n)}\mathds{1} - {\mathbf{H}})\mathbf{c}^{(n)} \delta\tau - \Delta\check{\mathbf{H}}\mathbf{c}^{(n)} ,
\end{align}
where  the fluctuating matrix $\Delta\check{\mathbf{H}}$ has zero mean and the matrix elements are given by
\begin{align} \label{eq:FCIQMCWienerH}
\Delta\check{{H}}_{ij} &= \sqrt{|S^{(n)}\delta_{ij} - {H}_{ij}|} \Delta \check{W}_{ij} .
\end{align}
This representation reproduces the mean and variances of the sampling procedure while approximating  Bernoulli distributed random numbers by normally distributed Wiener increments as explained in more detail in  App.~\ref{sec:sampling}. It presents an excellent starting point for further analysis.

We can now proceed to derive an It\^o differential equation for the walker number by norm-projection and taking the differential equation limit. The procedure is analogous  to the derivation of the scalar model in Sec.~\ref{sec:scalar} and is written out in detail in App.~\ref{sec:normSDE}. Using a variable transformation and It\^o's lemma to decorrelate the fluctuating shift from the walker number yields the following expression for the
difference between the shift and the norm projected energy estimators:
\begin{align} \label{eq:pcblowdensity_main}
\langle S\rangle - \bar{E}_{\tilde{\mathbf{1}}} &= \left\langle \frac{1}{2 N_\mathrm{w}^2}   \sum_{i,j} \left| S\delta_{ij} - {H}_{ij} \right|  c_j\right\rangle .
\end{align}
Comparing with Eq.~\eqref{eq:nEnergy} the above yields
an explicit expression for the covariance of shift and walker number.
Equation \eqref{eq:pcblowdensity_main} should be compared with the corresponding expression \eqref{eq:scalarPCB2} from the scalar model. In contrast to the scalar model, where the right hand side simply evaluated to a  constant of the model, we have a different situation here, where products of the fluctuating quantities of shift $S$, walker number $N_\mathrm{w}$, and state vector $\mathbf{c}$ appear explicitly.

Note that the right hand side of Eq.~\eqref{eq:pcblowdensity_main} depends on the fluctuating walker number $N_\mathrm{w}(t)$ as well as on the individual coefficient vector elements $c_j(t)$. While the latter may be expected to scale proportional to the norm  $N_\mathrm{w}$, which would lead to an overall  $N_\mathrm{w}^{-1}$ scaling, this is not necessarily the case in the sparse walker regime for which this equation was derived. However, we have observed $N_\mathrm{w}^{-1}$ scaling numerically even in the case of non-universal scaling of the population control bias, as shown in the inset of Fig.~\ref{fig:normprojectedenergy}.


\subsection{Application to Bose-Hubbard chain} \label{sec:BHM}

In order to understand more about the scaling properties of the right hand side of Eq.\ \eqref{eq:pcblowdensity_main}, we need to know something about the Hamiltonian and about where the state vector is probing it.
Thus it is plausible that the result will depend on the physics of the problem.
The sum goes over all non-zero coefficients $c_j$ and over all off-diagonal matrix elements that connect to it.
We thus specialize in the following to the Bose Hubbard Hamiltonian with $N$ particles in $M$ sites of Eq.\ \eqref{eq:bhm}.

\subsubsection{Low density superfluid}
When the bosons are well separated in a mostly empty lattice, then each of them can hop left or right, contributing $2J$ to the energy. The diagonal contribution is proportional to the shift, since there is no interaction energy in this regime (we consider the limit where $U$ is small and negligible).

The bias term in Eq.~\eqref{eq:fciqmcSDE} that we obtained from applying It\^o's lemma can  be simplified to
\begin{align}
\frac{1}{2 N_\mathrm{w}^2}   \sum_{i,j} \left| S\delta_{ij} -{H}_{ij} \right| c_j = \frac{|S|}{2  N_\mathrm{w}} + \frac{J N}{ N_\mathrm{w}}  .
\end{align}
Equation \eqref{eq:pcblowdensity} for the difference between average shift and projected energy  thus becomes
\begin{align} \label{eq:lowDpcb}
\langle S\rangle -  \langle E_{{\mathbf{1}}}\rangle  &
=  \left\langle \frac{J N + \frac{1}{2}|S| }{ N_\mathrm{w}}\right\rangle ,
\end{align}
where $N$ is the number of bosons. The diagonal contribution is proportional to the shift, since there is no interaction energy in this regime. One problem with this expression is that we still have products of fluctuating and correlated quantities inside the averages and thus we cannot rigorously separate them into a product of averages. However, under the assumption that the variable transformation to the logarithmic walker number has already given us the leading contribution to the covariance between shift and walker number, we may hope that the remaining covariances are of smaller order of magnitude. Taking only the leading terms  we approximate  $\langle E_{{\mathbf{1}}}\rangle \approx \bar{E}_{\tilde{\mathbf{1}}}$ and obtain
\begin{align} \label{eq:biasU0N}
\frac{\langle S\rangle}{N} -  \frac{\bar{E}_{\tilde{\mathbf{1}}}}{N}
 &\approx  \frac{J  + \frac{1}{2}\langle |S|/N \rangle}{ \left\langle N_\mathrm{w} \right\rangle} ,
\end{align}
where we have divided by particle number in order to relate to the intensive energy per particle. It is seen that the right hand side is approximately independent of particle number except for a small particle number dependence that could appear due to fluctuations in the shift.

\begin{figure}
\includegraphics[width=\columnwidth]{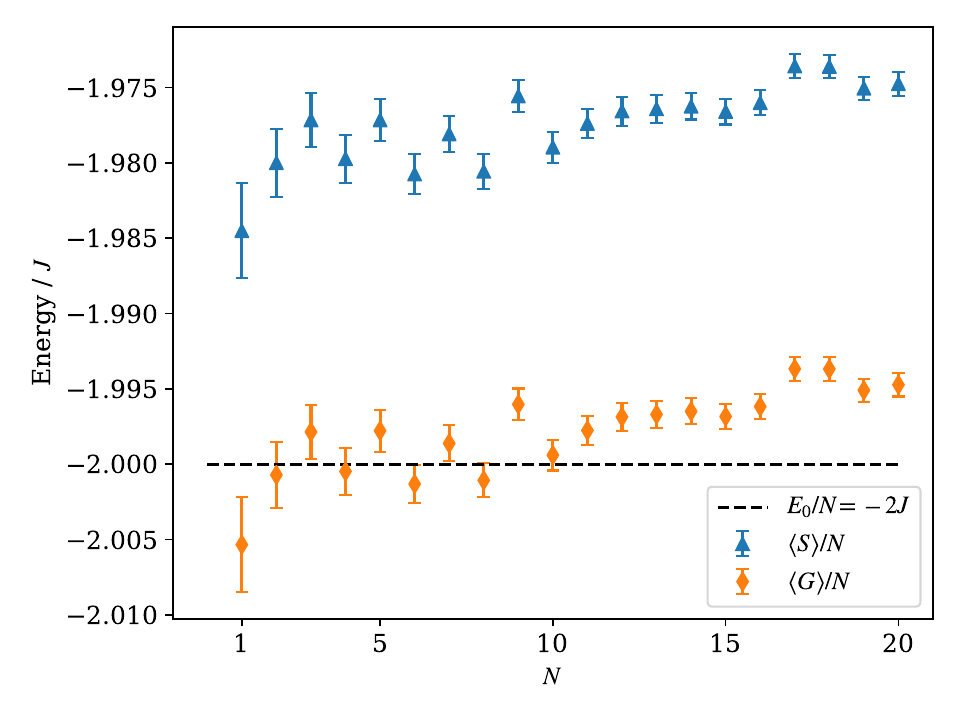}
 \caption{Shift estimator for the energy per particle in the low density superfluid regime ($U=0$) as a function of particle number $N$ in a lattice with $M=50$ lattice sites with $N_\mathrm{t} = 100$ walkers. The dashed black line shows the value of the exact energy per particle $E_0/N =  -2J$.
The deviation of the data from the exact results represents the population control bias, which is seen to be weakly dependent on the particle number $N$ as predicted by Eq.~\eqref{eq:biasU0N}.}
 \label{fig:NdependenceU0}
\end{figure}

Figure \ref{fig:NdependenceU0} shows the shift estimator and the projected energy estimator approximated by the growth estimator $\langle G\rangle$ as a function of particle number $N$ in the low density superfluid regime. It is seen that the difference in the energy estimators per particle is indeed nearly independent of $N$ as suggested by Eq.~\eqref{eq:biasU0N}. Furthermore, the projected energy estimator appears to have very little remaining population control bias for this system.


\subsubsection{Single particle Hubbard}

Further simplifications are found if we take $N=1$, where a very simple exact solution is known. In this case the Hilbert space is $M$ dimensional and spanned by the configurations $\hat{a}_i^\dag |\mathrm{vac}\rangle$. The (unnormalized) ground state is given by
\begin{align}
|\Psi_0\rangle = \sum_{i=1}^M \hat{a}_i^\dag |\mathrm{vac}\rangle ,
\end{align}
and thus the coefficient vector $\mathbf{c}_0$ of the exact ground state is a vector of all ones, i.e.~it is identical to the vector of all ones  ${\mathbf{1}}$ that we previously used to obtain the one-norm by projection
\begin{align}
\mathbf{c}_0 = {\mathbf{1}}.
\end{align}
The ground state further has the eigenvalue $E_0 = -2J$.
The considerations of the previous section still apply, with the difference that the norm projected energy estimator now becomes the exact (non-fluctuating) ground state energy
\begin{align}
 E_{{\mathbf{1}}} = \bar{E}_{\tilde{\mathbf{1}}} = E_0 = -2J .
\end{align}
We thus obtain an expression for the full population control bias from Eq.~\eqref{eq:lowDpcb}
\begin{align} \label{eq:PCB1particle}
\langle S\rangle -  E_0 &
=  \left\langle \frac{J  + \frac{1}{2}|S| }{ N_\mathrm{w}}\right\rangle ,
\end{align}
which is exact except for the assumptions made around It\^o calculus of Gaussian noise elements and an underlying Wiener process in continuous time.
Approximating this expression further by simply replacing the fluctuating walker number $ N_\mathrm{w}$ by $N_\mathrm{t}$ and using $S=-|S|$ and $E_0 = -2J$, we obtain
\begin{align} \label{eq:pcbSingleParticle}
\langle S\rangle -  E_0 &\approx  \frac{2J }{N_\mathrm{t}} .
\end{align}
as the leading term in $1 / N_\mathrm{t}$.

\begin{figure}
\includegraphics[width=\columnwidth]{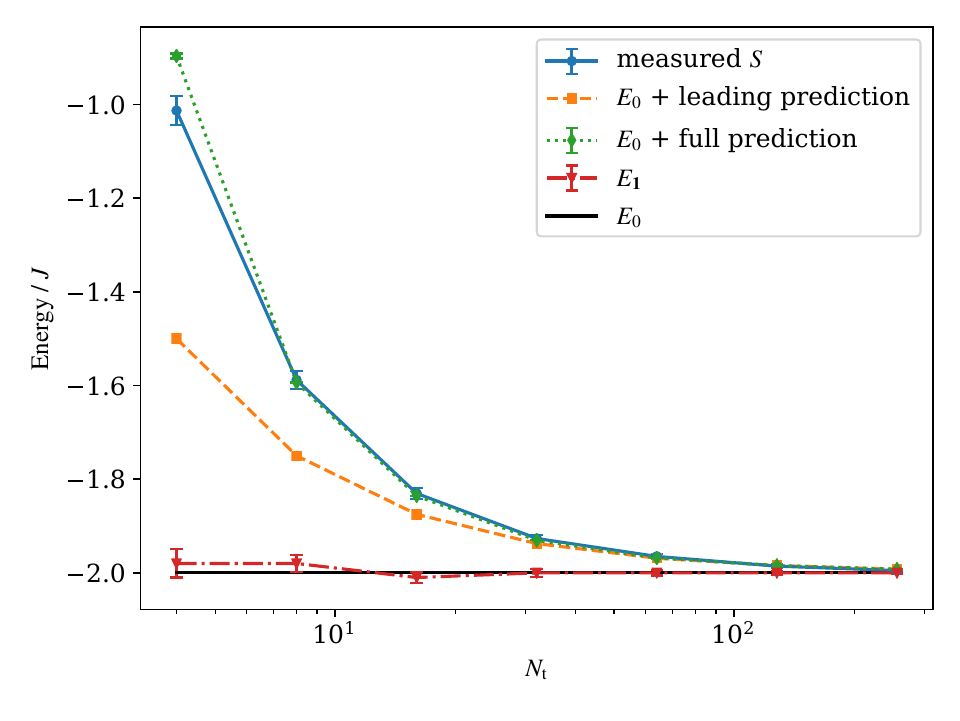}
 \caption{Shift estimator vs.~walker number for the single particle Hubbard model (blue circles) compared to the prediction of Eq.~\eqref{eq:PCB1particle} (green diamonds) and the leading power law from Eq.~\eqref{eq:pcbSingleParticle} (orange squares). The exact energy is shown as the full (black) horizontal line, and the red triangles corresponds to the measured shift corrected by the covariance term, i.e. $\bar{E}_{\tilde{\mathbf{1}}}$ as per Eq.~\eqref{eq:nEnergy}.
 } \label{fig:sp-lin}
\end{figure}
\begin{figure}
\includegraphics[width=\columnwidth]{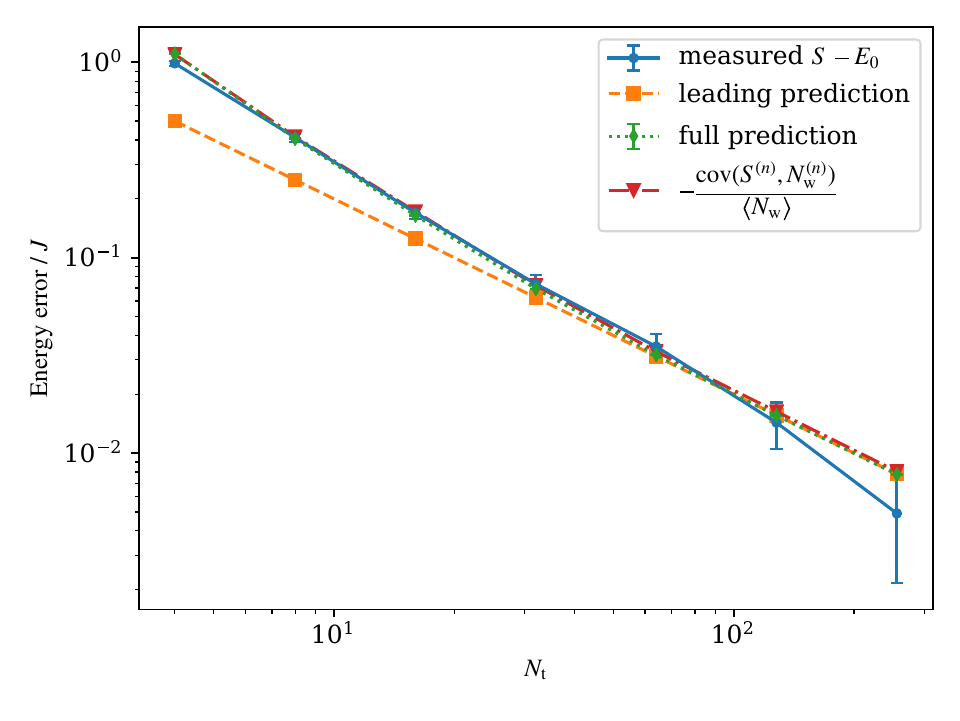}
 \caption{Energy error vs.~walker number on doubly logarithmic scales with the same data as in Fig.~\ref{fig:sp-lin}. Note that the measured shift estimator is consistent with the pure power law from Eq.~\eqref{eq:pcbSingleParticle} (orange squares, leading prediction) for large walker numbers, but decays more rapidly for small walker numbers. } \label{fig:sp-log}
\end{figure}

Figures \ref{fig:sp-lin} and \ref{fig:sp-log} demonstrate that the full prediction of Eq.~\eqref{eq:PCB1particle} works very well but the leading-order $1/N_\mathrm{w}$ prediction shows a significant discrepancy at small walker number. This is attributed to the fact that the shift was fluctuating across zero with significant amplitude and thus the assumption that the sign is consistently negative was not satisfied.


%

%
%
%

\section{\label{sec:conclusion}Conclusion}


In this work we have analysed the FCIQMC algorithm using tools of stochastic calculus.
This enabled us to derive bounds for the various estimators and find explicit solutions for the correlation functions in the time series of the walker number and the shift. The scalar model explains in particular why the population control bias
is independent of many parameters of the simulation. For the  damping and forcing parameters of walker control this independence was already seen previously in numerical data \cite{Yang2020}. Further empirical data demonstrating also the independence from the time step size and delayed update parameters is presented in Appendix \ref{sec:params}.


Our derivations of Sec.~\ref{sec:exactpcb} further showed that the shift estimator is an upper bound for the exact energy and other estimators, like the projected energy. This was derived for sign--problem free Hamiltonians and should also be true above the annihilation plateau for general Hamiltonians. It further provides a justification for the heuristic rule that the shift and projected energy estimators should agree when the population control bias has successfully been controlled.

We were also able to derive exact expressions for the population control bias for the very simple Hamiltonian of a single particle in the Hubbard chain. While the norm projected energy is an unbiased estimator in this case, the population control bias for the shift asymptotically scales with the inverse walker number $N_\mathrm{w}^{-1}$. While the $N_\mathrm{w}^{-1}$ scaling is consistent with previous works that have argued for this to be a universal feature of projection Monte Carlo methods  \cite{Hetherington1984,Umrigar1993,Cerf1995,Lim2015}, it is remarkable that we found non-universal scaling with slower power laws for Mott insulating states with particle numbers larger than about 20 in the Bose Hubbard model. We have verified numerically that the covariance of shift and walker number (and thus the difference between the shift and projected norm estimator) scale with  $N_\mathrm{w}^{-1}$, which indicates that the non-universal, slow power law scaling affects both of the energy estimators equally and may have a separate origin from the shift-walker number correlations. Our results were obtained with the original integer-walker sampling procedure of Ref.~\cite{Booth2009} and it remains an open question whether other sampling procedures like semistochastic and non-integer FCIQMC \cite{Petruzielo2012,Blunt2015} or fast randomized iteration algorithms \cite{Lim2015,Greene2019,Greene2020} would exhibit the same non-universal behavior.

An important question, naturally, is how the population control bias can be avoided, or mitigated. 
The reweighting procedure \cite{Nightingale1986,Nightingale1988,Umrigar1993}  discussed in Sec.~\ref{sec:Reweighting} is an interesting option, which can remove the bias in existing Monte Carlo time series in post-processing. While successful in removing small biases, it comes at the cost of increased stochastic errors. These can lead to signal loss when the bias is large, and  determining the optimal reweighting depth is difficult. 

Reducing the sampling noise is an obvious strategy that will reduce the bias in the time series along with stochastic errors. Importance sampling has already been demonstrated to achieve this \cite{Inack2018a}, and can be combined with reweighting \cite{Ghanem2021}.
The effectiveness for reducing the population control bias of other noise reduction strategies like semistochastic FCIQMC  \cite{Petruzielo2012,Blunt2015}, fast randomized iteration \cite{Lim2015,Greene2019,Greene2020}, and heat-bath sampling \cite{Holmes2016} remains to be assessed.
For strongly correlated problems in large Hilbert spaces, however, sampling noise cannot be fully avoided. 
The insight obtained from It\^o's lemma is that  the squared amplitude of the sampling noise finds its way back into the average of the shift, and thus causes the population control bias. An intriguing possibility is the option to inject additional imaginary noise, whose squared amplitude provides a negative contribution and can thus compensate the bias caused by the original sampling noise. We will concentrate future work in this direction and explore noise compensation with complex walker populations.

%

\begin{acknowledgments}
We thank Ali Alavi, Cyrus Umrigar, and Sebastiano Pilati for helpful discussions, and Matija \v{C}ufar for discussion and code improvements in particular with respect to replica calculations.
This work
was supported by the Marsden Fund of New Zealand (Contract No.\ MAU1604), from government funding managed by
the Royal Society of New Zealand Te Ap\={a}rangi. We also acknowledge support by the New Zealand eScience Infrastructure (NeSI) high-performance computing facilities in the form of a merit project allocation and a software consultancy project.
\end{acknowledgments}

\section*{Note added}

After the bulk of this work was completed we became aware of a recent preprint on the population control bias in FCIQMC by Ghanem \emph{et al.}~\cite{Ghanem2021}. While the results mostly complement our work, there is some overlap with  Sec.~\ref{sec:exactpcb}. We note that Ref.~\cite{Ghanem2021} argues for an $N_\mathrm{w}^{-1}$ scaling of the FCIQMC population control bias, while we provide counter examples demonstrating non-universal scaling in Sec.~\ref{sec:evidence}.
Reference \cite{Ghanem2021} further present data on two model systems where the bias could be removed by  combining noise reduction through importance sampling with reweighting of Monte Carlo data in post-processing (see Sec.~\ref{sec:Reweighting}, which was added later, and Refs.~\cite{Nightingale1986,Umrigar1993,Vigor2015}).


\appendix

\section{Dependence of the bias on other simulation parameters} \label{sec:params}

The FCIQMC equations \eqref{eq:Cupdate} and \eqref{eq:Supdate} contain  the time step $\delta\tau$, the damping constant $\zeta$, and the forcing $\xi$ as parameters of the simulation. Some formulations of FCIQMC delay updating the shift by $A$ steps, introducing an additional parameter. 
%
One may wonder  how these parameters influence the values or the biases of the energy estimators considered in this work. We have not seen any significant dependence  of the energy estimators on either of these parameters in our simulation, and present some exemplary evidence for the absence of such a parameter dependence in this section.

\begin{figure}
\includegraphics[width=\columnwidth]{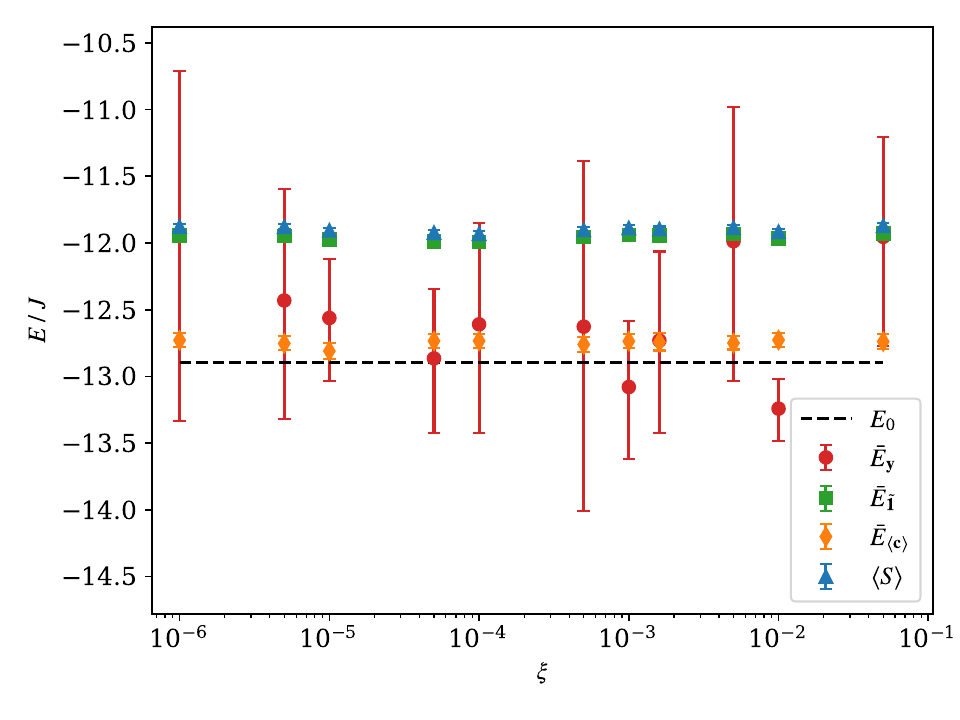}
\caption{Bias under forced population control.
Biased energy estimators against the forcing parameter $\xi$ from the real-space Bose-Hubbard chain with $N=M=20$. The projected energy estimator $\bar{E}_\mathbf{y}$ (red circles) with a projector containing the non-interacting ground state and all 40 connected (singly-excited) configurations shows much larger error bars than the other energy estimators. All estimators including the shift $\langle S\rangle$ (blue triangles), the norm-projected energy $\bar{E}_{\tilde{\mathbf{1}}}$ from Eq.~\eqref{eq:pEcov} (green squares), and the variational energy $\bar{E}_{\langle\mathbf{c}\rangle}$ (orange diamonds) show no significant dependence on the forcing parameter $\xi$ within the error bars. This implies that the forced population control of Eq.~\eqref{eq:Supdate} introduced in Ref.~\cite{Yang2020} is neutral with respect to the population control bias compared to the original procedure of Ref.~\cite{Booth2009}, where $\xi=0$.
The calculation was performed with $N_\mathrm{t} = 10^3$ walkers and $4\times 10^6$ time steps after equilibration.
The reference ground-state energy $E_0 = 12.90 J$ was obtained from an accurate calculation with $N_\mathrm{t}=10^7$ walkers.
All other parameters are chosen as in Fig.~\ref{fig:variationalenergy}.
}
\label{fig:estimatorswithxi}
\end{figure}

Figure \ref{fig:estimatorswithxi} shows various energy estimators while the forcing parameter $\xi$ is varied over several orders of magnitude. The limit $\xi=0$ corresponds to the unforced population control of Ref.~\cite{Booth2009} used in most of the literature to date, and $\xi =\zeta^2/4 = 0.0016$ is the value used in Fig.~\ref{fig:variationalenergy} and recommended in Ref.~\cite{Yang2020}. The absence of any significant $\xi$ dependence of the energy estimators indicates that the population control bias for all energy estimators is not affected by the population control mechanism introduced in Ref.~\cite{Yang2020}. Note that the population control bias in the shift estimator was already reported for a smaller system in Fig.~10 of Ref.~\cite{Yang2020}.

\begin{figure}
\includegraphics[width=\columnwidth]{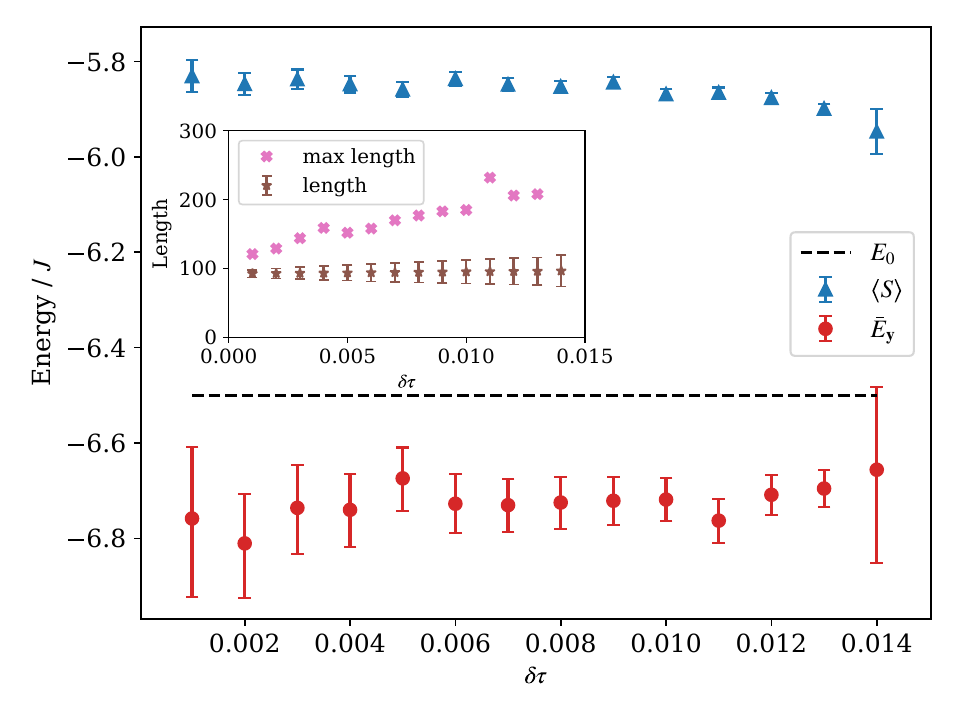}
\caption{Bias and stability as a function of the time step parameter.
Shift (blue triangles) and projected energy (red circles) vs.\ the size of the time step $\delta\tau$ for a Bose-Hubbard chain with $N=M=10$ and $U/J=6$. The exact energy (dashed line) was obtained from exact diagonalization. The bias in both energy estimators is evident and largely independent of $\delta\tau$. Stable simulation data was obtained for time steps beyond the theoretical stability threshold of $\delta\tau\approx 0.0072J^{-1}$ (see text) up to the largest value shown $\delta\tau = 0.014J^{-1}$, which is marginally unstable.
Due to the fixed number of time steps taken ($2^{20}$), the error bars decrease with increasing time step size. The inset shows the mean (stars), standard deviation (error bars) and the maximum (crosses) of the number of nonzero elements in the coefficient vector during the simulation.
Other parameters: $N_\mathrm{t}=100$, $\zeta = 0.08$, $\xi = \zeta^2/4$.
}
\label{fig:biasdeltatau}
\end{figure}

It is well known [and easy to derive from Eq.~\eqref{eq:Cupdate}] that the deterministic FCIQMC propagator has no time step error and is stable as long as $\delta\tau<2/(E_\mathrm{max} - E_0)$, where $E_\mathrm{max}$ and $E_0$ are the largest and smallest eigenvalue of the Hamiltonian matrix \cite{Spencer2012}. As seen in Fig.~\ref{fig:biasdeltatau} we find that also the population control bias in both the shift and the projected energy estimator is independent of the time step parameter $\delta\tau$.
It can be clearly seen that the error bars are decreasing with increased $\delta\tau$, indicating that the simulation is more efficient for larger time steps. As a trade-off, the number of non-zero elements in the coefficient vector fluctuates more, as seen in the inset of Fig.~\ref{fig:biasdeltatau}. For $\delta\tau \gtrsim 0.014 J^{-1}$ we see rapid, uncontrolled growth of the walker number and non-zero vector elements consistent with the instability of the FCIQMC equations. For the specific Hamiltonian of Fig.~\ref{fig:biasdeltatau} the stability boundary is  $2/(E_\mathrm{max} - E_0) \approx 0.0072 J^{-1}$, so smaller by a factor of two compared to the observed value of the instability.

\begin{figure}
\includegraphics[width=\columnwidth]{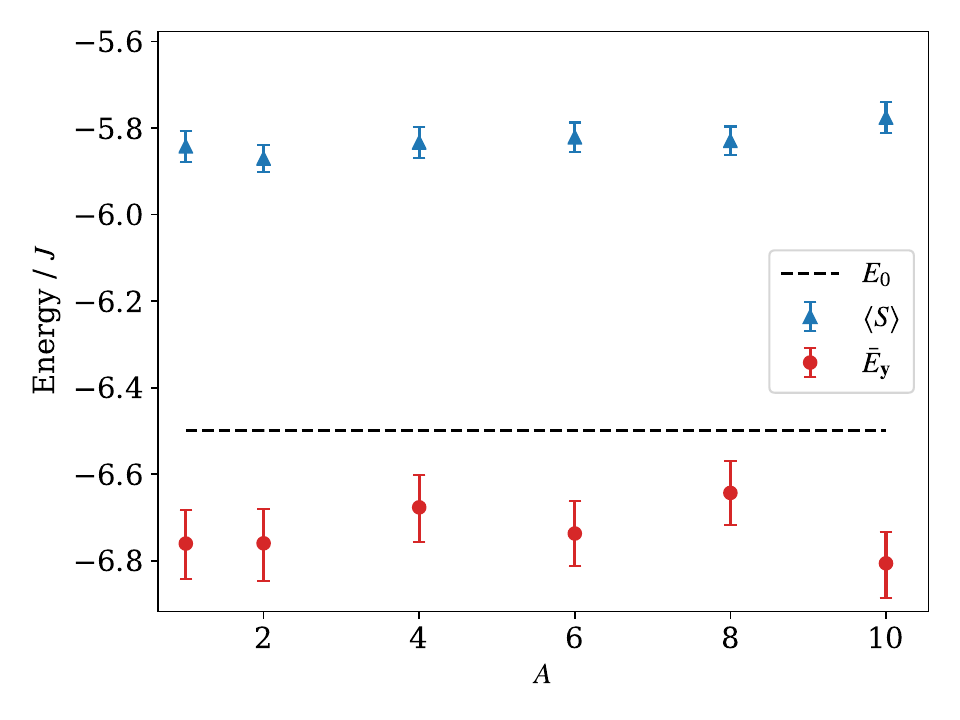}
\caption{Bias under delayed shift updates.
Shift (blue triangles) and projected energy (red circles)  vs.\ the delay $A$ in updating the shift parameter in comparison with the exact energy (dashed line). 
The model and parameters are as in Fig.~\ref{fig:biasdeltatau} and the time step is $\delta\tau=0.001$. 
Both energy estimators show a population control bias, but no significant dependence on the parameter $A$ is observed.
}
\label{fig:pcbwitha}
\end{figure}

In the original formulation of FCIQMC \cite{Booth2009} it was suggested to update the shift parameter not in every time step but rather only every 5 to 10 time steps. One might wonder whether delaying the shift updates influences the population control bias, as the fluctuating nature of the shift can be understood as the origin of the bias.
Delayed shift updates can be achieved by generalizing the shift update equation \eqref{eq:Supdate} to 
\begin{align}
S^{(n+A)} & = S^{(n)}- \frac{\zeta}{A\delta\tau}\ln\frac{N_\mathrm{w}^{(n+A)}}{N_\mathrm{w}^{(n)}}
- \frac{\xi}{A\delta\tau}\ln\frac{N_\mathrm{w}^{(n+A)}}{N_\mathrm{t}} ,
\end{align}
where  $A\ge1$ is the delay. In Ref.~\cite{Yang2020} we already examined in detail the effect of increasing $A$  and found that   the variance of the shift decreases while  the variance of the walker number increases. Importantly, the standard error of the shift estimator after blocking analysis, which is a measure for the statistical efficiency, was found unaffected by the parameter $A$. 

In Fig.~\ref{fig:pcbwitha} we show the values of the shift and projected energy estimators as a function of $A$ in an example calculation where the population control bias is appreciable. No significant dependence of the energies (or the bias) on $A$ is detected. This can be rationalized by the fact that the decrease of the variance of the shift for increasing $A$ is accompanied by increased fluctuations in the coefficient vector, and thus is not effective in suppressing the covariance responsible for the bias as per Eq.~\eqref{eq:pcb}. Furthermore, the population control bias is captured in the differential equation limit of the scalar model in Sec.~\ref{sec:SDElimit} where the parameter $A$ disappears and thus becomes irrelevant when the limit of an infinitesimal time step is taken. As setting $A>1$ has the disadvantage of larger fluctuations in the walker number and thus increased storage requirements for the coefficient vector while it brings no advantages, we recommend setting $A=1$, i.e.~retaining Eq.~\eqref{eq:Supdate} for performing shift updates.

The demonstrated independence of the population control bias of the values of the parameters $\delta\tau$, $\zeta$,  $\xi$, and $A$
is expected and supported by the theoretical arguments of Secs.~\ref{sec:scalar} and \ref{sec:noiseinfciqmc} where these parameters drop out of the final expressions for the population control bias. The empirical results of this appendix thus provide further support to the validity of our model. In  addition, we found empirically that FCIQMC simulations can be stable for values of the time step parameter $\delta \tau$ almost twice  the theoretical bound for instability.

\section{Shift estimator as upper bound for projected energy} \label{sec:upperbound}

Here we show that
\begin{align} \label{eq:ineqEy}
\left\langle S \right\rangle  - \bar{E}_\mathbf{y} \ge 0 ,
\end{align}
i.e.~the inequality of Eq.~\eqref{eq:pEcov}, which means that the shift energy estimator is greater or equal than any projected energy estimator. The proof follows a very similar logic and procedure as the derivation of the scalar model for the walker number in Sec.~\ref{sec:scalar} using It\^o's lemma. However, instead of using projection onto the $\tilde{\mathbf{1}}$ vector, we project onto an arbitrary vector $\mathbf{y}$. The only important assumption is that overlap $\mathbf{y}^\dag\mathbf{c}^{(n)}$ of the vector $\mathbf{y}$ with the coefficient vector $\mathbf{c}^{(n)}$ is non-zero at every time step.

Projecting the FCIQMC equation \eqref{eq:Cupdate} for the coefficient vector from the left with $\mathbf{y}^\dag$ yields an equation for the scalar projection
\begin{align} \label{eq:yupdate}
y^{(n+1)} - y^{(n)} & =  \mathbf{y}^\dag \delta\tau(S^{(n)}\mathbf{c}^{(n)}  - \check{\mathbf{H}}\mathbf{c}^{(n)}) ,
\end{align}
where we have introduced the notation
\begin{align}
y^{(n)} = \mathbf{y}^\dag\mathbf{c}^{(n)} ,
\end{align}
for the projection of the instantaneous vector.
The right hand side of Eq.~\eqref{eq:yupdate} is just the change in the coefficient vector during a single FCIQMC time step. We may replace it by its ensemble average (over an ensemble of random numbers in the sampling process), and a remaining noise term
\begin{align}
 \mathbf{y}^\dag \delta\tau(S^{(n)}\mathbf{c}^{(n)}  - \check{\mathbf{H}}\mathbf{c}^{(n)})  =
  \delta\tau (S^{(n)}- \bar{E}_\mathbf{y}) y^{(n)} +   r^{(n)} \Delta\check{W}^{(n)} .
\end{align}
Since the noise term has a variance proportional to $\delta\tau$ (see arguments in Secs.~\ref{sec:scalardiffeq} and \ref{sec:noiseinfciqmc}) and has to ensemble average to zero, we have written it as a product of a Wiener increment $\Delta\check{W}^{(n)}$, and a (still possibly fluctuating) factor $r^{(n)}$. The Wiener increment can be associated with a Wiener process $\check{W}(t)$ in continuous time $t$. Following the logic of Sec.~\ref{sec:SDElimit} we interpret the difference equation \eqref{eq:yupdate} as the Euler-Mayurama discretization of the It\^o stochastic differential equation
\begin{align}
dy =  [S(t)- \bar{E}_\mathbf{y}] y(t) dt +   r(t) d\check{W}(t) .
\end{align}
We aim at isolating the long-time average of the shift variable $S(t)$, which, however here sits in a product with the fluctuating variable $y(t)$. The latter can be removed by a variable transformation to
\begin{align}
z(t) = \ln y(t).
\end{align}
The variable transformation has to be performed according to It\^o's lemma \cite{enwiki:1008274165}. Using
\begin{align}
dz = \frac{1}{y}dy -\frac{1}{2y^2} dy^2 ,
\end{align}
together with the It\^o rules \eqref{eq:ItoRules} yields
\begin{align}
dz = \left[S(t) - \bar{E}_\mathbf{y} -\frac{r(t)^2}{2 y(t)^2}\right] dt + \frac{r(t)}{y(t)}  d\check{W}(t) .
\end{align}
In this form, a long time average can be taken term by term. In the steady state limit, the change in $z$  averages to zero, $\langle dz \rangle =0$, as does the average containing the Wiener noise term
\begin{align}
\left\langle \frac{r(t)}{y(t)}  d\check{W}(t) \right\rangle = 0 .
\end{align}
Collecting the remaining terms we obtain
\begin{align}
S(t) - \bar{E}_\mathbf{y} = \frac{r(t)^2}{2 y(t)^2} ,
\end{align}
where the right hand side is non-negative due to being a product of squares of real numbers. This completes the proof of inequality \eqref{eq:ineqEy} .

\section{Solutions of the SDEs with Greens functions}
\label{sec:GreensFunctions}

This appendix details the derivation of solutions of the SDEs \eqref{eq:scalarSDE1} and \eqref{eq:scalarSDE2}, obtaining expressions for the evolution of the logarithmic walker number $x(t)$ and the shift $S(t)$ for the cases of critical and non-critical damping.
For this purpose, we introduce a vectorized notation
\begin{align}
\mathbf{u}(t) = \binom{x(t)}{S(t)}.
\end{align}
This leads to Eqs.~\eqref{eq:scalarSDE1} and \eqref{eq:scalarSDE2} being re-written as
\begin{align} \label{eq:inhom_ode}
\mathbf{A} d\mathbf{u} + \mathbf{B} \mathbf{u}(t) dt = d\mathbf{f}(t) ,
\end{align}
where
 \begin{align}
\mathbf{A} = & \begin{pmatrix}
 1 & 0 \\
 \frac{\zeta}{d \tau} & 1
\end{pmatrix}, \\
\mathbf{B} = & \begin{pmatrix}
 0 & -1 \\
 \frac{\xi}{d \tau^2} & 0
\end{pmatrix}, \\
d\mathbf{f}(t) = &\binom{-(\tilde{E} +\frac{1}{2}\mu^2) dt - \mu dW(t)}{0} .
\end{align}
In order to solve the inhomogeneous linear differential equation \eqref{eq:inhom_ode} we seek a matrix-valued Greens function
\begin{align}
\mathbf{G}(t) = \begin{pmatrix}
 g_{11}(t) & g_{12}(t) \\
 g_{21}(t) & g_{22}(t)
\end{pmatrix},
\end{align}
that solves
\begin{align} \label{eq:GFtime}
\mathbf{A} \frac{d \mathbf{G}(t -t')}{dt} + \mathbf{B} \mathbf{G}(t -t') = \delta(t-t') \mathds{1} ,
\end{align}
where $\mathds{1}$ is the $2\times 2$ unit matrix.
Then
\begin{align} \label{eq:GFsolution}
\mathbf{u}(t) = \mathbf{u}_\mathrm{h}(t) +  \int_{-\infty}^{+\infty} \mathbf{G}(t-t') d\mathbf{f}(t') ,
\end{align}
is the general solution of the differential equation \eqref{eq:inhom_ode} where $\mathbf{u}_\mathrm{h}(t)$ is a solution of the corresponding homogeneous equation.
\begin{align} \label{eq:hom_ode}
\mathbf{A} d\mathbf{u} + \mathbf{B} \mathbf{u}(t) dt = 0 .
\end{align}
Being interested in the fluctuating steady-state solution, we take $x_\mathrm{h}(t) = 0 = S_\mathrm{h}(t)$ as the homogeneous solution. From this we readily obtain Eqs.~\eqref{eq:x_sol} and \eqref{eq:s_sol}.

In order to find the correct Greens function, we go into the frequency domain by Fourier transformation.
Defining the Fourier transform by
\begin{align}
\mathbf{G}(t) = \int_{-\infty}^{\infty} \frac{d\omega}{2\pi} e^{-i \omega t} \widetilde{\mathbf{G}}(\omega) ,
\end{align}
we can write Eq.~\eqref{eq:GFtime} in the frequency domain as
\begin{align}
-i\omega \mathbf{A}  \widetilde{\mathbf{G}}(\omega) + \mathbf{B}  \widetilde{\mathbf{G}}(\omega) = \mathds{1} ,
\end{align}
and solve for the Greens function as
\begin{align}
 \widetilde{\mathbf{G}}(\omega) &= (-i\omega \mathbf{A} +\mathbf{B})^{-1} ,\\
 &= \frac{1}{(\omega - \omega_-)(\omega - \omega_+)}
 \begin{pmatrix}
 i\omega & -1 \\
 -i\omega\frac{\zeta}{\delta\tau}+\frac{\xi}{\delta\tau^2} &  i\omega
\end{pmatrix} ,
\end{align}
where
\begin{align}
\omega_{\pm} &=-i\frac{\zeta}{2\delta\tau}\pm i \sqrt{ \frac{\zeta^2}{4\delta\tau^2} - \frac{\xi}{ \delta\tau^2}} , \\
& = -i\gamma \pm i \tilde{\gamma} ,
\end{align}
are two complex poles corresponding to the two different damping coefficients, or frequencies, of the damped harmonic oscillator solution.
The Greens function in the time domain is obtained by Fourier transformation
\begin{align} \label{eq:gtime}
\mathbf{G}(t) & =  \int_{-\infty}^{\infty} \frac{d\omega}{2\pi}  \frac{e^{-i \omega t}}{(\omega - \omega_-)(\omega - \omega_+)}
 \begin{pmatrix}
 i\omega & -1 \\
 -i\omega\frac{\zeta}{\delta\tau}+\frac{\xi}{\delta\tau^2} &  i\omega
\end{pmatrix}.
\end{align}
The integral can be solved by contour integration after closing the contour in the upper or lower complex half plane using Jordan's lemma depending on the sign of $t$. Accordingly, the contour integral either evaluates to zero (when no poles are enclosed), or is given by the sum of the residues of the enclosed poles.

\subsection{Greens function for critical damping}
The special case $\omega_+ = \omega_- \equiv -i\gamma = -i\frac{\zeta}{2\delta\tau} = -i\frac{\sqrt{\xi}}{\delta\tau}$ corresponds to critical damping of the harmonic oscillator. Here the residue theorem gives
\begin{align}
\mathbf{G}(t) &= i  \theta(t)
\lim_{\omega \to -i\gamma} -\frac{d}{d\omega} e^{-i \omega t} \begin{pmatrix}
 i\omega & -1 \\
 - 2i\omega \gamma +\gamma^2&  i\omega
\end{pmatrix} \\
&=  \theta(t)  \begin{pmatrix}
1-\gamma t & t \\
-2\gamma + \gamma^2 t & 1-\gamma t
\end{pmatrix} e^{-\gamma t} ,
\end{align}
where  the Heaviside function $\theta(t) = 1$ when $t>0$ and $\theta(t) = 0$ otherwise.
From the first column we obtain the explicit expression of Eqs.~\eqref{eq:g11_crit} and \eqref{eq:g21_crit} for $g_{11}(t)$ and $g_{21}(t)$.
%

\subsection{Greens function for over- and under-damped case}
In the more general case, we have $\omega_+ \neq \omega_-$ and the integrand of Eq.~\eqref{eq:gtime} has simple poles. The integral evaluates to
\begin{align}
\nonumber
\mathbf{G}(t) &= \theta(t) \sum_{\sigma \in \{-,+\}} \frac{i \sigma e^{-i \omega_\sigma t}}{\omega_+ - \omega_-} \\
&\quad \times
 \begin{pmatrix}
 i\omega_\sigma & -1 \\
 -i\omega_\sigma\frac{\zeta}{\delta\tau}+\frac{\xi}{\delta\tau^2} &  i\omega_\sigma
\end{pmatrix} .
\end{align}
Specifically
\begin{align}
g_{11}(t) &= \theta(t) e^{-\gamma t} \left[ \frac{\gamma}{\tilde{\gamma}} \sinh(\tilde{\gamma}t) - \cosh(\tilde{\gamma}t)\right] , \\
g_{21}(t) &= \theta(t) e^{-\gamma t} \left[\frac{\zeta}{\delta\tau} \cosh(\tilde{\gamma}t) +\frac{\xi-\frac{1}{2}\zeta}{\delta\tau^2 \tilde{\gamma}} \sinh(\tilde{\gamma}t)\right] .
\end{align}
For the over-damped case $\tilde{\gamma} =  \sqrt{ \frac{\zeta^2}{4\delta\tau^2} - \frac{\xi}{ \delta\tau^2}}$ is real-valued while for the under-damped case it is purely imaginary, which serves to replace hyperbolic by trigonometric functions.


\section{Sampling noise in sparse walker regime} \label{sec:sampling}

We specifically investigate the algorithm with integer walkers of Ref.~\cite{Booth2009}, although the general logic should apply to non-integer spawning with threshold and similar sampling algorithms as well. We further simplify the analysis by assuming that we are in the low walker density regime (i.e. the walker number is much smaller than the Hilbert space dimension) where we have at most a single walker on each configuration. This is the regime where the effect of stochastic noise will be the largest and the population control bias will be the most severe.
Let's rewrite Eq.~\eqref{eq:Cupdate} to separate the spawning and diagonal death/cloning steps
\begin{align} \label{eq:CDOD}
\mathbf{c}^{(n+1)} -  \mathbf{c}^{(n)}  & = \delta\tau(S^{(n)} - \check{\mathbf{H}}_\mathrm{D})\mathbf{c}^{(n)} - \delta\tau  \check{\mathbf{H}}_\mathrm{OD}\mathbf{c}^{(n)} ,
\end{align}
where $\check{\mathbf{H}} = \check{\mathbf{H}}_\mathrm{D}+\check{\mathbf{H}}_\mathrm{OD}$ separates the fluctuating Hamiltonian into a diagonal matrix and a purely off-diagonal matrix. First we consider noise in the off-diagonal part, which relates to the spawning process, before turning to the diagonal part.

\subsection{Off-diagonal sampling: Spawning noise}

In a low density limit we assume that the elements of the coefficient vector ${c}^{(n)}_i$  only ever take the values $0$ or $\pm 1$.
A single spawning attempt corresponds to evaluating (at $i\neq j$)
\begin{align} \label{eq:randspawn}
-\delta\tau  \check{H}_{ij}  {c}^{(n)}_j &=
\begin{cases}
\pm 1 & \textrm{if }\, \mathtt{rand}  < \left|\delta\tau {H}_{ij}  {c}^{(n)}_j\right| ,\\
0 & \textrm{else} ,
\end{cases}
\end{align}
where $\mathtt{rand} \in [0,1)$ is a uniformly drawn random number and the sign is carried consistently. This characterizes the spawning of a single walker and defines a random variable following a Bernoulli distribution. 
We can now evaluate the expectation value
\begin{align} \label{eq:exp_spawn}
\left\langle-\delta\tau  \check{{H}}_{ij}  {c}^{(n)}_j  \right\rangle_\mathrm{e} & = -\delta\tau {H}_{ij}  {c}^{(n)}_j ,
\end{align}
where $\left\langle \cdot  \right\rangle_\mathrm{e}$ denotes an ensemble expectation value according to the random numbers drawn in each spawning event, while ${c}^{(n)}_j$ is just a given number. For the long-time averages considered elsewhere in this work, these coefficients are considered fluctuating quantities.
Since the expression \eqref{eq:exp_spawn} only ever can take values of zero or $\pm 1$, the expectation value of the squared expression can also be easily evaluated
\begin{align}
\left\langle\left(-\delta\tau  \check{{H}}_{ij}  {c}^{(n)}_j \right)^2 \right\rangle_\mathrm{e} & = \left| \delta\tau {H}_{ij}  {c}^{(n)}_j \right| .
\end{align}
For the standard deviation $\sigma_\mathrm{e}(x) = \sqrt{\langle x^2\rangle_\mathrm{e} - \langle x\rangle_\mathrm{e}^2}$ we thus obtain
\begin{align}
\sigma_\mathrm{e}\left(-\delta\tau  \check{{H}}_{ij}  {c}^{(n)}_j  \right) & = \sqrt{ \left| \delta\tau {H}_{ij}  {c}^{(n)}_j \right| -   \left( \delta\tau {H}_{ij}  {c}^{(n)}_j \right)^2} ,\\
& \approx  \sqrt{ \left| \delta\tau {H}_{ij}  {c}^{(n)}_j \right|} , \label{eq:approxsigma}
\end{align}
where we have used the fact that the squared factor is much smaller than 1, which can be assured by considering the limit of small $\delta\tau$.

It is thus justified to treat the randomness in the FCIQMC spawning process by a matrix $\check{\mathbf{H}}_\mathrm{OD}$ with random elements $ \check{{H}}_{ij}$ which is characterized by an expectation value
\begin{align}
\left\langle  \check{{H}}_{ij} \right\rangle_\mathrm{e} = {H}_{ij} ,
\end{align}
and standard deviation as above.

It is convenient to write
\begin{align} \label{eq:introWiener}
-\delta\tau  \check{{H}}_{ij} {c}^{(n)}_j   = -\delta\tau {H}_{ij}{c}^{(n)}_j  + \sqrt{ \left| {H}_{ij} \right|} \Delta \check{W}_{ij} {c}^{(n)}_j  ,
\end{align}
where the coefficient $ {c}^{(n)}_j$ could be pulled out of the square root because its value is either 0 or $\pm 1$ and thus $\sqrt{ \left|  {c}^{(n)}_j \right| }=  \left|  {c}^{(n)}_j \right|$. The right hand side of Eq.~\eqref{eq:introWiener} has the correct expectation value and standard deviation [in the approximation \eqref{eq:approxsigma}] if
\begin{align}
\langle \Delta \check{W}_{ij}\rangle_\mathrm{e} & = 0 , \\
\left\langle (\Delta \check{W}_{ij})^2 \right\rangle_\mathrm{e} & = \delta\tau .
\end{align}
In the following we will identify $\Delta \check{W}_{ij}$ with a Wiener increment, i.e.\ a Gaussian random variable characterized by its expectation value and variance as above. This is not exactly true because the spawning process is not Gaussian  but instead follows a Bernoulli distribution, and thus higher moments of the elementary process will differ. However, we will eventually be adding the effects of many spawning events, which will lead to a binomial distribution and approximate a normal distribution.Thus the Gaussian approximation at the single event level may not be such a bad one. Note that the random numbers $\Delta \check{W}_{ij}$ for different indices $(i,j)$ are independent and are also freshly drawn for each time step.

\subsection{Diagonal death noise}

The diagonal death step can be treated in a similar fashion as the spawning process in the previous subsection. It relates to evaluating the first term on the right hand side of Eq.~\eqref{eq:CDOD}. A single diagonal death attempt corresponds to evaluating
\begin{align}
\delta\tau (S^{(n)} -  \check{{H}}_{jj})  {c}^{(n)}_j &=
\begin{cases}
\pm 1 & \textrm{if }\, \mathtt{rand} < p_d ,\\
0 & \textrm{else} ,
\end{cases}
\end{align}
where the $p_d = |\delta\tau (S^{(n)} -  {H}_{jj})  {c}^{(n)}_j|$ is the death probability. The sign is carried consistently such that
\begin{align}
\left\langle \delta\tau (S^{(n)} -  \check{{H}}_{jj})  {c}^{(n)}_j  \right\rangle_\mathrm{e} & =\delta\tau (S - {H}_{jj})  {c}^{(n)}_j ,\\
\left\langle\left[ \delta\tau (S^{(n)} -  \check{{H}}_{jj})  {c}^{(n)}_j  \right]^2 \right\rangle_\mathrm{e} & = \left| \delta\tau (S^{(n)} -  {H}_{jj})  {c}^{(n)}_j \right| .
\end{align}
We can now follow the logic of the previous subsection to write the diagonal death step as the sum of a deterministic process, and a random process with expectation value zero that is approximated with a Wiener increment:
\begin{align} \label{eq:diagnoise}
\nonumber
\delta\tau (S^{(n)} -  \check{{H}}_{jj})  {c}^{(n)}_j = & \, (S^{(n)} -  {H}_{jj})  {c}^{(n)}_j  \delta\tau + \\
&+ \sqrt{| S^{(n)} -  {H}_{jj}|} \Delta \check{W}_{jj} {c}^{(n)}_j .
\end{align}

Combining the diagonal and off-diagonal processes of Eqs.~\eqref{eq:diagnoise} and \eqref{eq:introWiener}, respectively, to a matrix equation yields Eq.~\eqref{eq:FCIQMCWiener}.
%
%

\section{Stochastic differential equation for FCIQMC walker number} \label{sec:normSDE}

In this appendix we derive an It\^o SDE for the FCIQMC walker number and shift in the sparse walker regime starting from the representation of the coefficient update Eqs.~\eqref{eq:FCIQMCWiener} and \eqref{eq:FCIQMCWienerH}.

We proceed by norm projection similar to Sec.~\ref{sec:scalardiffeq}, with the difference that we are  keeping track of the individual noisy matrix elements for now. It is further convenient to specialize
to a  \emph{stoquastic} Hamiltonian  and  assume that all coefficients $c_i$ are non-negative, i.e.~either have value 0 or 1 in the sparse walker regime. In this case we can obtain an equation for the walker number (or one-norm) by projecting the vector valued equation on the vector of all ones $\mathbf{1}$:
\begin{align} \label{eq:CWiener}
\nonumber
\Vert\mathbf{c}^{(n+1)}\Vert_1 - \Vert\mathbf{c}^{(n)} \Vert_1 = & (S^{(n)}\Vert\mathbf{c}^{(n)} \Vert_1 - {\mathbf{1}}^\dagger \mathbf{H} \mathbf{c}^{(n)} )
\delta\tau + \\
&+{\mathbf{1}}^\dagger\Delta \check{\mathbf{H}} \mathbf{c}^{(n)} .
\end{align}
Let us write this as a differential equation and use the notation $ \Vert\mathbf{c}^{(n)} \Vert_1 \equiv N_\mathrm{w}^{(n)} \to N_\mathrm{w}(t)$ for the norm as previously
\begin{align}
dN_\mathrm{w} = (S N_\mathrm{w} -   {\mathbf{1}}^\dagger \mathbf{H} \mathbf{c} ) \, dt + {\mathbf{1}}^\dagger d\check{\mathbf{H}} \mathbf{c},
\end{align}
where the last term represents a linear combination of many Wiener noises. In order to eliminate the product of fluctuating variables $Sc$, we perform a variable transformation to $x(N_\mathrm{w}) = \ln (N_\mathrm{w}/N_\mathrm{t})$ and use It\^o's lemma with Eq.\ \eqref{eq:changeofvars}
\begin{align}
dx & = \left(S - E_{{\mathbf{1}}} \right)\, dt - \frac{1}{2 N_\mathrm{w}^2} \left( {\mathbf{1}}^\dagger d\check{\mathbf{H}} \mathbf{c}\right)^2
+ \frac{{\mathbf{1}}^\dagger d\check{\mathbf{H}} \mathbf{c}}{N_\mathrm{w}} ,
\end{align}
where
\begin{align}
E_{{\mathbf{1}}} = \frac{ {\mathbf{1}}^\dagger \mathbf{H} \mathbf{c}}{N_\mathrm{w}} ,
\end{align}
is a (fluctuating) projected energy. The noise term evaluates to
\begin{align} \label{eq:noiseTerm}
{\mathbf{1}}^\dagger d\check{\mathbf{H}} \mathbf{c} &= \sum_{ij} \sqrt{|S \delta_{ij} - {H}_{ij}|}  d\check{W}_{ij} c_j .
\end{align}
For the squared noise term we obtain
\begin{widetext}
\begin{align}
 \left( {\mathbf{1}}^\dagger d\check{\mathbf{H}} \mathbf{c}\right)^2 &= \sum_{i,j,i',j'}\sqrt{ \left| (S\delta_{ij} - {H}_{ij})(S\delta_{i'j'} - {H}_{i'j'})\right|} \,d\check{W}_{ij}  d\check{W}_{i'j'} c_j c_{j'} , \\
 &= \sum_{i,j} \left| S\delta_{ij} - {H}_{ij} \right| c_j\,  dt ,
\end{align}
\end{widetext}
because $dW_{ij}  dW_{i'j'} = dt\, \delta_{ii'} \delta_{jj'}$  according to It\^o rules. We have also replaced the $c_j^2$ by $c_j$ consistent with the low-walker density limit.
The It\^o SDE for $x(t)$ then finally takes the form
\begin{align} \label{eq:fciqmcSDE}
dx & = \left(S - E_{{\mathbf{1}}}- \frac{1}{2 N_\mathrm{w}^2}   \sum_{i,j} \left| S\delta_{ij} -{H}_{ij} \right| c_j\right)\, dt
+ \frac{{\mathbf{1}}^\dagger d\check{\mathbf{H}} \mathbf{c}}{N_\mathrm{w}} .
\end{align}
In the steady-state regime, the statistical average of $dx$ on the left, and the noise term on the right, vanish. Thus we obtain
\begin{align} \label{eq:pcblowdensity}
\langle S\rangle - \langle E_{{\mathbf{1}}}\rangle &= \left\langle \frac{1}{2 N_\mathrm{w}^2}   \sum_{i,j} \left| S\delta_{ij} - {H}_{ij} \right|  c_j\right\rangle ,
\end{align}
which is an approximate expression for the population control bias.
Note that $\langle E_{{\mathbf{1}}}\rangle = \langle G\rangle =  \bar{E}_{\tilde{\mathbf{1}}}$ are all equivalent expressions for the norm projected energy within It\^o calculus with infinitesimal time step. This concludes the derivation of Eq.~\eqref{eq:pcblowdensity_main}.


\bibliography{shift-update,Methods}

\end{document}